\documentclass[aps,prb,superscriptaddress,amsfonts,amsmath,amssymb]{revtex4}
\usepackage{bm}
\usepackage{graphicx}
\usepackage{amsmath}
\usepackage{amstext}
\usepackage{amssymb}
\usepackage{amsfonts}
\usepackage{amsbsy}
\usepackage{verbatim}

\def\be{\begin{equation}}
\def\ee{\end{equation}}
\def\ba#1{\begin{array}{#1}}
\def\ea{\end{array}}
\def\bn{\begin{enumerate}}
\def\en{\end{enumerate}}

\def\r{\right}
\def\l{\left}

\def\summ{\sum\limits}
\def\o{\omega}

\def\s{{K_{s}}}

\def\sg{{K^g_s}}
\def\vf{\varphi}
\def\vfk{\tilde{\varphi}}
\def\fk{\tilde{\phi}}

\def\intt{\int\limits}
\def\tc{J}

\def\br{\boldsymbol{r}}
\def\bR{\boldsymbol{R}}
\def\bk{\boldsymbol{k}}
\def\bK{\boldsymbol{K}}

\def\on{\omega_{n}}
\def\kmeas{\frac{d^2 \bk}{(2 \pi)^2}}

\begin{document}

\title{Universal point contact resistance between thin-film superconductors}
\author{Michael Hermele}
\affiliation{Department of Physics, Massachusetts Institute of Technology, Cambridge, Massachusetts 02139, USA}
\author{Gil Refael}
\affiliation{Department of Physics, California Institute of Technology, Pasadena, California 91125, USA}
\author{Matthew P. A. Fisher}
\affiliation{Kavli Institute for Theoretical Physics, University of
California, Santa Barbara, California 93106, USA}
\author{Paul M. Goldbart}
\affiliation{Department of Physics, University of Illinois at
Urbana-Champaign, 1110 West Green Street, Urbana, Illinois 61801-3080, USA}

\date{\today}
\begin{abstract}
A system comprising two superconducting thin films connected by a
point contact is considered.  The contact resistance
is calculated as a function of temperature and film geometry, and is found to 
vanish rapidly with temperature,
according to a universal, nearly activated form, becoming strictly
zero only at zero temperature.  At the lowest temperatures, the
activation barrier is set primarily by the superfluid stiffness in
the films, and displays only a weak (\emph{i.e.}~logarithmic)
temperature dependence.  The Josephson effect is thus destroyed,
albeit only weakly, as a consequence of the power-law-correlated
superconducting fluctuations present in the films below the
Berezinskii-Kosterlitz-Thouless transition temperature.  The
behavior of the resistance is discussed, both in various limiting
regimes and as it crosses over between these regimes.  Details are presented of a minimal model of the films and
the contact, and of the calculation of the resistance.
A formulation
in terms of quantum phase-slip events is employed, which is natural
and effective in the limit of a good contact.  However, it is also
shown to be effective even when the contact is poor and is, indeed,
indispensable, as the system always behaves as if it were in the good-contact limit at
low enough temperature.
A simple mechanical analogy is introduced to provide
some heuristic understanding of the nearly-activated
temperature dependence of the resistance.  
Prospects for experimental tests of the
predicted behavior are discussed, and numerical estimates relevant
to anticipated experimental settings are provided.
\end{abstract}
\maketitle

\section{Introduction}
\label{sec:intro}

The Josephson effect\cite{josephson62}
 is one of the most direct probes of superconducting long-range order (LRO).
Phase coherence -- and hence a dissipationless supercurrent -- can be maintained across a variety of weak links between two bulk three-dimensional (3d) superconductors.  While one can imagine the decay of a supercurrent across the link via some dissipative process such as a phase slip, this would require the phase to ``unwind'' deep into the bulk superconductors, which is prevented by the rigidity of the order parameter.

In low-dimensional systems, fluctuations are more important and lead to a host of remarkable phenomena.  Superconductors are no exception to the rule, and a variety of low-dimensional or quasi-low-dimensional systems where superconducting fluctuations play an important role are the subject of much current interest.  For example, several experiments have investigated a so-called superconductor-insulator transition in
superconducting nanowires.\cite{bezryadin00,lau01,tinkham02,bollinger04} These
experiments measured the resistance of thin wires connecting two superconducting electrodes.
Theoretical work on the subject
\cite{zaikin97,golubev01,refael03} has explored the possibility of universal length- and width-dependent behavior of such devices. 
Also, superconducting phase fluctuations play an important role in some strongly correlated and disordered systems.  Noteworthy examples include the pseudogap regime of the cuprate superconductors\cite{emery95} and  disordered two-dimensional (2d) films near a superconductor-insulator transition.\cite{haviland89, hebard90, gantmakher98}

A striking aspect of low-dimensional systems is that the fluctuations can destroy true long-range phase order, but sometimes only weakly so.  Rather than a resistive state with only short-range superconducting correlations, one finds power-law superconducting fluctuations, \emph{i.e.} quasi-long-range order (QLRO), and vanishing bulk resistivity.  This is the case in one-dimensional (1d) superconducting wires at zero temperature, and in 2d films below the Berezinskii-Kosterlitz-Thouless (BKT) phase transition at temperature $T_{{\rm BKT}}$.

What is the fate of the Josephson effect for such low-dimensional superconductors?  
The one-dimensional case of this question, together with a host of related issues, has been the source of much experimental and theoretical interest.\cite{kane92-prl,kane92,chang96,egger97,kane97,grayson98,furusaki98,bockrath99,auslaender00}  While this work has been pursued in the context of various physical systems, the common thread is that the 1d conducting ground state is a Luttinger liquid with power-law superconducting correlations and zero resistivity.  The simplest system consists of two 1d wires joined by a point contact.  There, provided the superfluid stiffness of the wires is greater than a certain critical value, the tunneling resistance vanishes as a power-law in temperature, with a universal exponent depending only on the stiffness.\cite{kane92-prl,kane92}  This power law is a reflection of the critical nature of the zero-temperature Luttinger-liquid ground state of the wires.

\begin{figure}
\includegraphics[width=8cm]{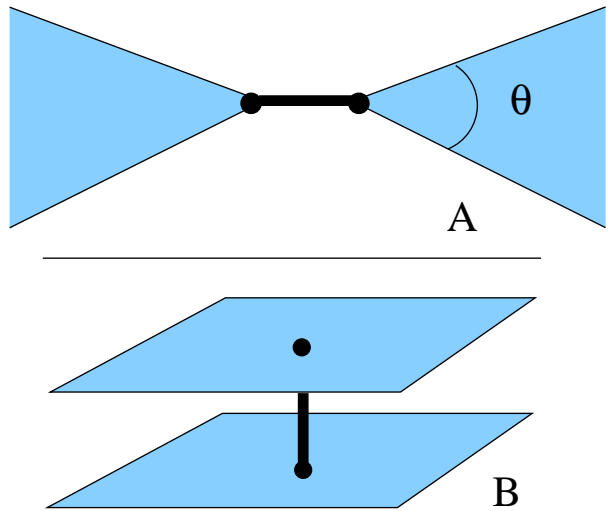}
\caption{We consider both wedge (A) and bilayer (B) geometries. The shaded regions are the superconducting films, and the dark line joining them represents the point contact.  Note the definition of the opening angle $\theta$ in the wedge geometry.}
\label{fig:geomfig}
\end{figure}

Perhaps the simplest system for studying Josephson physics in \emph{two-dimensional} superconductors consists of two superconducting thin films connected by a point contact, as illustrated schematically in Fig.~\ref{fig:geomfig}A.  The point contact can be any kind of weak link much smaller than the films themselves.
Systems of this kind, where two MoGe films are joined by a thin nanowire\cite{bezryadin00, lau01} or a single narrow constriction,\cite{chu04} have been the subject of recent transport experiments.  A related situation has been studied in
the experiment of Ref.~\onlinecite{naaman01}, where Pb films were probed by the tunneling of Cooper pairs from a superconducting STM tip.
Aside from the work of Kim and Wen,\cite{ybkim93} which we discuss further below,
very little theoretical attention has been paid to this problem.

Motivated by these experiments, and in anticipation of further work on similar systems, we have studied theoretically the d.c.~transport across a point contact between two thin-film superconductors.  A nontechnical discussion of our main results can be found in Ref.~\onlinecite{shortpaper}.  We focus on linear-response properties; for example, we consider only the resistance given by the voltage response to a very weak current bias.  The point contact is characterized by the Josephson coupling energy $\tc$, or, equivalently, the critical current $I_c = 2 e \tc / \hbar$.  The films have superfluid stiffness $\s = \hbar^2 n_s / m$, where $n_s$ is the number of Cooper pairs per unit area of the film, and $m$ is the pair mass.  
The problem can be approached from the opposite limits of a poor contact ($\tc / \s$ small) or a good contact ($\tc / \s$ large).  The transport in the poor-contact limit is naturally viewed in terms of Cooper pair hopping events, where a single pair hops across the contact from one film to the other with amplitude $\tc$.  This limit was analyzed by Kim and Wen,\cite{ybkim93} who calculated the d.c.~conductance ${\cal G}(T)$ as a function of temperature, and found that it apparently diverged below a characteristic temperature $T^*$.\footnote{In fact, Ref.~\onlinecite{ybkim93} studied tunneling from a bulk 3d superconductor into a thin film.  Upon integrating out the modes away from the point contact, as is done here, that problem is easily seen to be formally equivalent to one of tunneling between two thin films.}

\begin{figure}
\includegraphics[width=8cm]{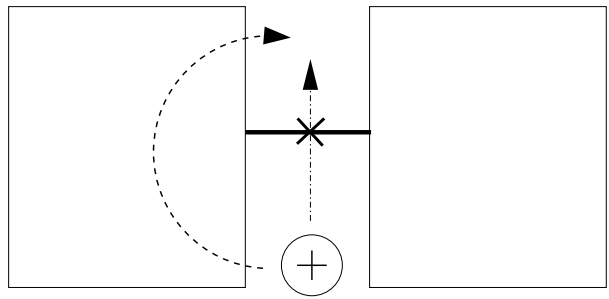}
\caption{Depiction of phase-slip processes.  In one, a vortex (ellipse with arrow) enters the left-hand superconducting film below the contact (dark line with ``X'' through it), follows the dashed-line path, and exits the film on the other side of the contact.  In the second, the vortex moves directly through the contact, along the dash-dot line.  This process corresponds to a continuous change of the phase at one end of the contact from zero to $\pm 2\pi$, while the phase at the other end remains fixed.}
\label{fig:vortex-hop}
\end{figure}

As we argue below, the result of Ref.~\onlinecite{ybkim93} is in error (except at zero temperature), and is symptomatic of a breakdown of perturbation theory rather than a physical zero-resistance state.  
We arrive at this conclusion by working in the \emph{good-contact} limit; in fact, one of our main results is that, at low enough temperature, the leading-order perturbative calculation in this limit is \emph{always} correct, even when $\tc \ll \s$.  Transport in the good-contact limit can be viewed in terms of \emph{quantum phase slips}; these are events where the order-parameter phase winds by $2\pi$ across the contact.  Phase slips correspond to vortex-hopping processes, in which a vortex enters one of the films near the contact and then moves to the other side of the contact, where it leaves the film (see Fig.~\ref{fig:vortex-hop}).  A different, but qualitatively identical, phase-slip process involves the phase at one end of the contact changing continuously in time from $0$ to $\pm 2\pi$; this can be visualized as a vortex hopping directly across the contact (Fig.~\ref{fig:vortex-hop}). The disturbance created by one of these events can heal locally, via the reverse process where a vortex hops in the opposite direction.  Alternatively, it can propagate out
into the films, causing a voltage spike across the system of films plus contact, according to the Josephson relation
\begin{equation}
\Delta V = \frac{\hbar}{2 e} \frac{d}{d t} \Delta \varphi \text{,}
\end{equation}
where $\varphi$ is the phase of the superconducting order parameter, and $\Delta V$ and $\Delta \varphi$ schematically represent the difference in voltage and $\varphi$, respectively, between two points. 
Formally, it is possible to work directly in terms of phase-slip events, via a duality transformation.  One can then calculate the resistance perturbatively in $t_v$, the amplitude for phase-slip events to occur.  

We note that the phase slips are quantum in nature because $t_v$ approaches a constant as the temperature is lowered to zero; in this limit phase slips are clearly quantum tunneling events.  Furthermore, we expect only a weak temperature variation in $t_v$ as long as $T \ll T_{{\rm BKT}}$.
However, the physics determining the low-temperature behavior of the resistance is that of \emph{classical} phase fluctuations within the superconducting films, except at strictly zero temperature.  
The disturbance caused by a vortex hopping across the contact is an excitation of the gapless plasmon mode in the films, and it is the physics of the plasmons that determines whether a voltage spike will register across the entire system.  Our detailed considerations show that the plasmon excitations important in determining the low-temperature resistance have frequency much lower than $k_B T / \hbar$, and are thus thermally populated (and hence in a classical regime) at any nonzero temperature.

We find that the resistance vanishes at $T = 0$, the only temperature where there is superconducting LRO in the films.  For all $T > 0$, on the other hand, the resistance is nonzero but vanishes following a universal form as $T \to 0$.  This implies that the perturbative poor-contact result is incorrect; at any fixed temperature, the resistance cannot decrease as the contact is weakened, so, as the resistance is nonzero in the good contact limit, a vanishing resistance is out of the question.  The low-temperature form is very nearly an activated exponential,
\begin{equation}
R(T) = R_Q \frac{\pi^{3/2} t_v^2}{4} \frac{1}{\sqrt{E_A(T) (k_B T)^3}}
\exp \Big( - \frac{E_A(T)}{k_B T} \Big) \text{,}
\label{eqn:resistance}
\end{equation}
where $R_Q \equiv h / 4 e^2$ is the quantum of resistance, and $E_A(T)$ is a temperature-dependent effective activation barrier:
\begin{equation}
E_A(T) = c \s \frac{1}{\ln(\hbar \omega_c / 2 k_B T) + (2 c \s / \pi^2 \bar{\tc})} \text{.}
\label{eqn:energy-barrier}
\end{equation}
Here, $\bar{\tc} \approx \tc$ is an effective Josephson coupling that need not be distinguished from $\tc$ for most purposes, and $\omega_c$ is a high-energy cutoff set by an appropriate short-distance cutoff in the films; these parameters are discussed in more detail below.
The superfluid stiffness $\s$ plays a crucial role in setting $E_A(T)$. The constant $c$ is a \emph{universal} dimensionless quantity given by
\begin{equation}
c = \frac{\pi^2 \theta}{4 \alpha} \text{,}
\label{eqn:defn-of-c}
\end{equation}
where $\theta$ is the ``opening angle'' at the point contact, as shown in Fig.~\ref{fig:geomfig}A.  For Coulomb interactions between electrons in the films we have $\alpha = 2$, whereas $\alpha = 1$ if the interactions are screened down to a short-range form. The latter case can be achieved in the presence of a superconducting ground plane separated from the films by an insulating layer [see Sec.~\ref{sec:ground-plane}]. 
 
As written, Eq.~(\ref{eqn:resistance}) is expected to be asymptotically exact as $T \to 0$, in the sense that $[\ln R(T)] / [\ln R_{{\rm measured}}] \to 1$ as $T \to 0$.   Furthermore, this formula should be a good approximation whenever $E_A(T) / k_B T \gtrsim 1$.  It is important to note that the ratio of the resistances themselves (as opposed to the ratio of their logarithms) will \emph{not} approach unity as $T \to 0$.  This is due to additive corrections to $E_A(T)$ not captured in Eq.~(\ref{eqn:energy-barrier}); the leading such correction is proportional to $1/ T \ln^2 T$.  These corrections, which are expected to depend on $\tc$, are recovered if the integrals leading to Eq.~(\ref{eqn:resistance}) are evaluated exactly.  This point  is discussed further in Secs.~\ref{sec:resistance-calculation} and \ref{sec:discussion}, where the integrals are written down in closed form.

Focusing on the barrier $E_A(T)$, the dependence on $\tc$ drops out at low temperature, where we find the \emph{universal} form:
\begin{equation}
E_A(T) \sim \frac{c \s}{\ln (\hbar \omega_c / 2 k_B T)} \text{.}
\label{eqn:universal-lowT-barrier}
\end{equation}
This form is universal in the sense that it does not depend on the details of the point contact; instead, the important temperature scale is set by $\s$, which is a property of the films themselves and can be determined independently by an inductance measurement.\cite{fiory83}  The same result holds in the limit of a good contact ($\tc \gg \s$) for \emph{all} $T \ll \s / k_B$.  At the lowest temperatures, then, the system always behaves as if it were in the good-contact limit.

On the other hand, when $\tc \ll \s$, \emph{provided the temperature is not too low}, we find that the barrier has the form
\begin{equation}
E_A(T) \sim \frac{\pi^2 \bar{\tc}}{2} \text{.}
\label{eqn:purely-activated-barrier}
\end{equation}
More precisely, for this to hold we require that $\ln( \hbar \omega_c / k_B T) \ll \ln ( \hbar \omega_c / k_B T_J)$, where $T_J$ is a crossover temperature defined by
\begin{equation}
T_J \equiv \frac{\hbar \omega_c}{2 k_B} \exp\Big( - \frac{2 c \s}{\pi^2 \bar{\tc}} \Big) \text{.}
\end{equation}
It should be noted that the condition to be in this limit involves the logarithm of the temperature, rather than the temperature itself.  In the opposite limit (\emph{i.e.}~low temperature), the resistance always crosses over to the universal form of Eq.~(\ref{eqn:universal-lowT-barrier}).
As shown in Appendix~\ref{app:integrals-poor-contact}, the form Eq.~(\ref{eqn:purely-activated-barrier}) is expected to be asymptotically exact when the above conditions (on $\tc / \s$ and the logarithm of the temperature) are satisfied, provided we also require that $\tc \gg k_B T$.

Therefore, Eq.~(\ref{eqn:energy-barrier}) is an approximate interpolation between two limits where it is exact.  However, the order-$t_v^2$ contribution to the resistance is expected to provide an essentially exact result provided $k_B T \ll \operatorname{min}(\s, \tc)$.  Our result for $E_A(T)$ comes from an approximate evaluation of this contribution, which is exact in the limits discussed above.  We provide a closed-form expression for the \emph{exact} ${\cal O}(t_v^2)$ contribution in Sec.~\ref{sec:resistance-calculation}, which can be evaluated numerically.  The reason that terms of higher-order in $t_v$ are not expected to be important is that these terms involve larger activation barriers, owing to the strong binding of phase slips in an imaginary-time functional-integral formulation; therefore, the ${\cal O}(t_v^2)$ term will dominate whenever $k_B T$ is smaller than $\s$ and $\tc$, as these energy scales set the barrier height.

The nearly-activated form of Eq.~(\ref{eqn:resistance}) is suggestive of the experimental results of Ref.~\onlinecite{chu04}.  Furthermore, as discussed in Sec.~\ref{sec:experiments}, it should be possible to carry out future experiments in a regime where a detailed comparison with our 
theory is appropriate.

\section{Outline of paper}

We continue in Sec.~\ref{sec:formulation} with a description of our model, which is a quantum phase Hamiltonian for the films that focuses on quantum and thermal fluctuations of the Cooper pair phase field.  The contact is modeled by a point-like Josephson coupling having strength $\tc$.  The degrees of freedom in the films are integrated out to obtain an effective action involving only the phase difference $\phi$ across the contact.  We also give some general arguments in favor of a treatment starting from the good-contact limit.

In Sec.~\ref{sec:coulomb-wedge} we discuss some subtleties in treating Coulomb interactions in the wedge geometry of Fig~\ref{fig:geomfig}A, and in Sec.~\ref{sec:ground-plane} we discuss a setup where the films and contact are placed in proximity to a superconducting ground plane, in order to screen the Coulomb interaction down to a short-range form.  In Sec.~\ref{sec:weak-coupling} we review the calculation of the conductance in the poor contact limit (\emph{i.e.} working perturbatively in powers of $\tc$).

In Sec.~\ref{sec:duality} we describe the duality transformation used to access the good-contact limit.  In Sec.~\ref{sec:resistance-calculation}, the resistance is calculated in this limit, to leading order in $t_v$ (\emph{i.e.}~the amplitude for quantum phase slips).  
Some further intuition for the physics behind the resistance formula is provided by a mechanical analogy, which we discuss in Sec.~\ref{sec:mechanical}.  

%Section~\ref{sec:variational} describes a variational approach, the results of which further support our %contention that the good contact limit is the correct starting point at low temperature.

In Sec.~\ref{sec:experiments} we discuss the prospects for experimental tests of our theory via
measurements of point-contact tunneling between two superconducting films.  We give necessary conditions for such systems to be in a regime where our theory applies.  We conclude in Sec.~\ref{sec:discussion} with a discussion of some of the open questions raised by our results for a variety of different systems.

Several appendices contain the more technical aspects of the work.
Appendix~\ref{app:actions} contains a derivation of the imaginary-time
action for the point contact, which is obtained by integrating out the
low-energy degrees of freedom in the films.  The analytic continuation
used to obtain the necessary correlation functions in real time is
discussed in Appendix~\ref{app:imag-time}.  The details of the
evaluation of the integrals used to obtain the resistance formula of
Eq.~(\ref{eqn:resistance}) are contained in
Appendix~\ref{app:integrals}. Appendix \ref{app:variational} presents
the details of a variational approach to the point-contact problem,
which provides another justification for the good-contact treatment of 
Sec.~\ref{sec:strongtcp}.

\section{Model of superconducting thin films connected by a point contact}
\label{sec:formulation}

We consider a system of two superconducting thin films connected by a point contact.  Essentially, this can be any weak link much smaller than the dimensions of the films.  In some cases, especially, for example, if the link is a superconducting wire, it is necessary to be sure that one is in a regime where any internal degrees of freedom of the link play no role and it can be treated as a zero-dimensional object.  We use spatial coordinates $\br$ defined separately for each film, so that the contact is at the origin $\br = \boldsymbol{0}$.
We are interested in the ``wedge'' geometry of Fig.~\ref{fig:geomfig}A, which is specified by the opening angle $\theta$.  Although we also consider the bilayer geometry of Fig.~\ref{fig:geomfig}B,  our treatment of that case is intended \emph{only} to illustrate features of the wedge setup in a technically simpler context.  In particular, we do not include the interlayer Coulomb interaction, which would be important in a real bilayer system.  However, it should be noted that a real bilayer system can be analyzed along the lines of Sec.~\ref{sec:ground-plane}, where we consider a system of films in the wedge geometry, separated from a superconducting ground plane by an insulating barrier.\cite{mh-unpub}  

Note that in this Section, as well as Secs.~\ref{sec:weak-coupling} through \ref{sec:mechanical} and the appendices, we shall set $\hbar = k_B = 1$.

We are interested in $T < T_{{\rm BKT}}$, which we assume to be substantially below the quasiparticle energy gap.  We can therefore focus on the Cooper-pair degrees of freedom in the superconducting films.  Within the superconducting state, the physics is encapsulated in a quantum phase Hamiltonian written in terms of the Cooper-pair phase $\varphi_i(\br)$ and density $n_i(\br)$, which satisfy the canonical commutation relations
$[\varphi_i(\br), n_j(\br')] = i \delta_{i j} \delta(\br - \br')$.  Here, $i, j = 1,2$ label the two films.  In the wedge geometry we must impose the condition that supercurrent cannot flow across the film edge; mathematically, this gives the boundary  condition $\nabla\varphi_i \cdot \boldsymbol{n} = 0$, where $\boldsymbol{n}$ is normal to the edge.  Due to this boundary condition, the case $\theta = 2 \pi$ does not correspond exactly to the bilayer geometry; however, we find identical results in these two cases.

The Hamiltonian for the system is
\begin{equation}
\label{eqn:full-hamiltonian}
{\cal H} = {\cal H}_1 + {\cal H}_2 - \tc \cos (\varphi_1(0) - \varphi_2(0)) \text{,}
\end{equation}
where the last term is the Josephson coupling at the contact, and ${\cal H}_{1,2}$ are the Hamiltonians for the thin films.  These take the form
\begin{equation}
\label{eqn:film-hamiltonian}
{\cal H}_i = \frac{\s}{2} \int d^2\br (\nabla \varphi_i)^2 + V_i \text{,}
\end{equation}
where $\s = \s(T)$ is the superfluid stiffness in units of energy, and $V_i$ is the interaction potential.  For Coulomb interactions this takes the usual form
\begin{equation}
\label{eqn:coulomb-potential}
V_i = \frac{(2 e)^2}{2} \int d^2\br d^2\br' \frac{n_i(\br) n_i(\br')}{| \br - \br' |} \text{.}
\end{equation}
We also consider the case of short-range interactions, which take the simple form:
\begin{equation}
\label{eqn:short-range-ints}
V_i = \frac{v_s^2}{2 \s} \int d^2\br \big(n_i(\br)\big)^2 \text{,}
\end{equation}
where $v_s$ is the superfluid velocity.  The short-range case directly describes a two-dimensional superfluid of charge-neutral bosons.  However, it can also be realized for a superconducting film if the Coulomb interaction is screened down to a short-range form.  This can be achieved in the presence of a superconducting ground plane, as discussed in Sec.~\ref{sec:ground-plane}.
In both cases we neglect inter-plane density-density interactions, which are expected to be unimportant in the wedge geometry.  Further justification for this expectation is given below.

The low-energy excitations in the films are the superconducting plasmons, whose thermal excitation is responsible for the QLRO occurring for $0 < T < T_{{\rm BKT}}$.  For screened interactions (and a translation-invariant film), one finds an acoustic plasmon having the linear dispersion $\omega = v_s |\bk|$.  In the 
case of Coulomb interactions the plasmon has the dispersion $\omega = 2 e \sqrt{2\pi \s |\bk|}$.

Note that we have completely neglected vortex fluctuations within the films.  We can justify this by estimating the typical number of vortices in a film of linear dimension $L$.  The energy cost of a single free vortex in a 2d superconductor is on the order of $\pi \s \ln(\Lambda / \xi)$ plus a core energy $E_{{\rm core}}$. Here, $\xi$ is the coherence length and $\Lambda = c^2 / 4\pi e^2 \s$ is the magnetic screening length.  $\Lambda$ can be quite large; for example, if $T_{{\rm BKT}} = 5\, {\rm K}$, $\Lambda(T_{{\rm BKT}}) \approx 0.8\, {\rm cm}$.
Now, the number of vortices in the whole film is approximately given by
\begin{equation}
N_v \approx \frac{L^2}{\xi^2} \Bigg(\frac{\xi}{\Lambda} \Bigg)^{\pi \s / T} \exp (-E_{{\rm core}} /  T )
% \nonumber \\
%&=& \frac{L^2}{\xi^2} \exp \Big( -\frac{E_{{\rm core}} + \pi \s \ln (\Lambda/\xi)}{k_B T} \Big) \text{.}
\end{equation}
At $T = T_{{\rm BKT}}$, the exponent $\pi \s / T$ takes on the universal value of 2, and, neglecting $E_{{\rm core}}$, $N_v \approx (L / \Lambda)^2$.  Thus, for $L = 1\, {\rm cm}$ films and $\Lambda = 0.8\,{\rm cm}$, $N_v$ is approximately unity at $T_{{\rm BKT}}$ and decreases rapidly with temperature.  Therefore, the 
typical separation between free vortices in the films is on the order of or greater than the system size $L$, and vortices can indeed be neglected.

%As shown in Fig.~\ref{fig:geomfig}, we discuss this problem in two geometries.  In the bilayer geometry (Fig.~\ref{fig:geomfig}A) both planes are infinite in all directions and full two-dimensional translation symmetry obtains in the limit of zero Josephson coupling.  In the wedge geometry (Fig.~\ref{fig:geomfig}B) the planes still have infinite extent but only within the region specified by the opening angle $\theta$.  

As the modes of the phase field in the films are governed by a harmonic theory, we can integrate them out to obtain an imaginary-time action for the phase difference across the contact, 
$\phi(\tau) \equiv \varphi_1({\bf r} = 0, \tau) - \varphi_2({\bf r} = 0, \tau)$.  Details of this procedure, which is straightforward except for the case of Coulomb interactions in the wedge geometry
(see Sec.~\ref{sec:coulomb-wedge}), are given in Appendix~\ref{app:actions}.  The result is $Z = \int{\cal D}\phi \exp(-S_0 - S_{J})$, where
\begin{equation}
\label{eqn:gaussian-phi-action}
S_0 = \frac{2 c \s T}{\pi^2} \sum_{\omega_n} \left[ \ln \frac{\omega_n^2 + \omega_c^2}{\omega_n^2} \right]^{-1} | \tilde{\phi}(\omega_n)|^2 \text{.}
\end{equation}
Here $c$ is defined in Eq.~(\ref{eqn:defn-of-c}),  $\omega_n \in 2\pi T {\mathbb Z}$, and we have defined the Fourier transform via
\begin{eqnarray}
\tilde{\phi}(\omega_n) &=& \int_0^{\beta} d\tau\, e^{i \omega_n \tau} \phi(\tau) \\
\phi(\tau) &=& T \sum_{\on} e^{-i \on \tau} \tilde{\phi}(\on) \text{,} 
\end{eqnarray}
where $\beta \equiv 1/T$.
Finally, $\omega_c$ is a high-frequency cutoff set by the short-distance cutoff in the films, which we take to be on the order of the coherence length $\xi$.  
For example, in the bilayer geometry $\omega_c = \omega(|\bk| = 2\pi/a_c)$.  
The remaining term in the action is simply the Josephson coupling across the contact:
\begin{equation}
\label{eqn:josephson-coupling}
S_J = - \tc \int_0^\beta d\tau\, \cos(\phi(\tau)) \text{.}
\end{equation}

We note that the summand of $S_0$ depends only weakly on $\on$; apart from the important logarithmic frequency dependence, it is simply a $|\tilde{\phi}(\on)|^2$ ``mass term'' for the $\phi$ field.  Such a term is obtained for a point contact between 3d superconducting electrodes, and is thus expected to lead to zero resistance even at finite temperature.  It is therefore apparent from the form of $S_0$ that the system of films plus contact is in a sense very close to exhibiting true superconducting behavior (\emph{i.e.}~a d.c.~Josephson effect at finite temperature).  It is not, however, clear \emph{a priori} whether the Josephson effect survives intact at finite temperature, or whether it is weakly destroyed.  Our analysis shows that the latter scenario is the correct one.

It is instructive to consider the physics at zero temperature, where we argue there will be no quantum phase transition as a function of $\tc$ or other parameters.  This can be seen by considering phase-slip instanton events in $\phi$, where $\phi \to \phi \pm 2\pi$ over some localized region in imaginary time.  It is useful to recall a different model,
 where $S_0$ is replaced by
 $S'_0 \propto \sum_{\on} | \on | |\tilde{\phi}|^2$.  In that case, very similarly to vortices in the 2d XY model, phase slips interact via a potential proportional to the logarithm of their separation, and there is a zero-temperature Chakravarty-Schmid phase transition as the coefficient of $S'_0$ is varied.\cite{chakravarty82, schmid83}  If the potential is stronger than a logarithm at asymptotically large times, the potential energy will always dominate over the entropy, and phase slips will be bound into ``neutral'' pairs (\emph{i.e.}~pairs of one $\phi \to \phi +2\pi$ phase slip and one $\phi \to \phi -2\pi$ phase slip).  This is exactly what happens in the present case, where we find a linearly confining potential (up to important logarithmic corrections).  Therefore, the system is always superconducting at $T=0$.
 
The very strong binding of phase slips already indicates that one should be cautious about the expansion in the poor contact limit.  There will be many events in imaginary time in which a Cooper pair tunnels across the contact, but the poor-contact expansion is most reliable when we can think of a dilute gas of such events.  On the other hand, the good-contact expansion, in which one thinks instead of a dilute gas of phase slips, should be very reliable.  

Some comments are now in order regarding our neglect of density-density interactions between the two films.  For films of finite size, inter-film interactions will lead to a Coulomb blockade below a temperature that vanishes as the film dimensions are taken to infinity.  This point is discussed further in Sec.~\ref{sec:experiments}; here we concentrate on the limit of infinite films.
For short-range interactions in the wedge geometry, such interactions are certainly unimportant. They will lead to an additive contribution to $S_0$ proportional to $\omega_n^2 | \tilde{\phi}|^2$.  At low frequency this term is dominated by the logarithmic term of Eq.~(\ref{eqn:gaussian-phi-action}) that came from the power-law phase fluctuations in the films, and there will be no effect on the low-temperature physics.  The case of inter-film Coulomb interactions is more subtle.  However, following Sec.~\ref{sec:ground-plane}, we have analyzed the bilayer geometry for films of equal stiffness, including the inter-film Coulomb interaction.\cite{mh-unpub}  This is a useful case to consider, because inter-film Coulomb interactions should have a stronger effect than in the wedge geometry.  It is convenient to exploit the symmetry under interchange of the films and make the change of variables $\varphi_{\pm}(\br) = (\varphi_1(\br) \pm \varphi_2(\br))/\sqrt{2}$, and similarly for the densities.  The Hamiltonian decouples into a sum of two terms, each depending only on either symmetric or antisymmetric fields.  The Josephson term couples only to the antisymmetric sector, so the symmetric sector does not contribute to the charge transport.  The main effect of the inter-film Coulomb interactions is to screen the interaction in the antisymmetric sector down to a short-range form, leading to a linearly-dispersing acoustic plasmon.  This is quite natural, as the density $n_-(\br)$ is that of charge-neutral dipoles of positively charged Cooper pairs in film one and their negatively charged counterparts in film two.  The qualitative structure of the effective action is completely unaffected, and, not surprisingly, one now finds $\alpha = 1$ as appropriate for a model with only short-range interactions.  
The effects of inter-film Coulomb interactions in the wedge geometry should be even less.  Indeed, we expect $\alpha = 2$ as if the inter-film interactions were absent.  This is because the special structure of the bilayer, which led via screening to a linearly-dispersing plasmon mode, is absent in this case, and the plasmons in both films should still follow $\omega \propto \sqrt{| \bk |}$ dispersions.

\subsection{Coulomb interactions in wedge geometry}
\label{sec:coulomb-wedge}

In the bilayer geometry it is completely straightforward to obtain the action Eq.~(\ref{eqn:gaussian-phi-action}) by working in Fourier space.  The problem is also simple for short-range interactions in the wedge geometry; the kinetic term in ${\cal H}_i$ is diagonal in a basis of eigenfunctions of the Laplacian,  and in this case $V_i$ is diagonal in any basis.  Coulomb interactions are more difficult to treat, as it is not straightforward to simultaneously diagonalize the kinetic and potential terms.  We believe it is possible to avoid this difficulty without affecting the result by replacing the Coulomb potential by a different form $V(\br, \br')$ that has some of the same properties, but is more amenable to analytic treatment.

In an infinite 2d plane, the Fourier transform of the Coulomb potential $V_{C}(\br) = 4 e^2/|\br|$ 
is 
\begin{equation}
\tilde{V}_{C}(\bk) = \int d^2\br\, e^{- i \bk \cdot \br} V_{C}(\br) = \frac{8\pi e^2}{|\bk|} \text{.}
\end{equation}
We can write  
\begin{equation}
V_{C} = 8\pi e^2 (\sqrt{- \nabla^2})^{-1} \text{,}
\label{eqn:sqrt-Laplacian}
\end{equation}
where
we define the square root of the Laplacian in a plane wave basis, that is
\begin{equation}
\sqrt{-\nabla^2} e^{i \bk \cdot \br} = |\bk| e^{i \bk \cdot \br} .
\end{equation}
We shall construct a potential in the wedge geometry for which the analog of Eq.~(\ref{eqn:sqrt-Laplacian}) is true.

In the wedge geometry it is natural to work in a basis of eigenfunctions of the Laplacian that respect the $\nabla \varphi \cdot \boldsymbol{n} = 0$ boundary condition.  In plane polar coordinates $(r,\psi)$, this condition takes the form
\begin{equation}
\partial_\psi \chi(r, \psi = 0) = \partial_\psi \chi(r, \psi = \theta) = 0.
\end{equation}
The eigenfunctions are labeled by a non-negative real number $\lambda$ and a non-negative integer $n$, and satisfy $-\nabla^2 \chi_{\lambda n} = \lambda^2 \chi_{\lambda n}$.  They take the form:
\begin{equation}
\chi_{\lambda n}(\br) = \left\{ \begin{array}{ll}
\sqrt{\frac{2 \lambda}{\theta}} J_{\pi n/\theta} (\lambda r) \cos(\frac{n \pi \psi}{\theta})\text{,} & n > 0  \text{,}\\
\sqrt{\frac{\lambda}{\theta}} J_0 (\lambda r) \text{,} & n = 0 \text{.}
\end{array} \right.
\label{eqn:eigenfunctions}
\end{equation}
Here, $J_\nu(x)$ is a Bessel function of the first kind.  These eigenfunctions are orthonormal,
\begin{equation}
\int_0^\infty dr\, r \int_0^\theta d\psi \, \chi_{\lambda n}(\br) \chi_{\lambda' n'}(\br)  = \delta_{n n'} \,\delta(\lambda - \lambda') \text{,}
\end{equation}
and satisfy a closure relation:
\begin{equation}
\int_0^\infty d\lambda \sum_{n=0}^{\infty} \chi_{\lambda n}(\br) \, \chi_{\lambda n}(\br') =
	\delta(\br - \br') \text{.}
\end{equation}

With this machinery at our disposal, we define $\sqrt{-\nabla^2}$ by:
\begin{equation}
\sqrt{-\nabla^2} \chi_{\lambda n} = \lambda \chi_{\lambda n} \text{.}
\end{equation}
This then leads to the potential:
\begin{equation}
V(\br, \br') = 8 \pi e^2 \int_0^\infty d\lambda \sum_{n=0}^\infty
\frac{\chi_{\lambda n}(\br) \, \chi_{\lambda n}(\br')}{\lambda} \text{,}
\label{eqn:new-potential}
\end{equation}
which satisfies $\sqrt{-\nabla_{\br}^2} V(\br, \br') = 8 \pi e^2 \delta(\br - \br')$. 
This function depends separately on $\br$ and $\br'$, and not only on $|\br - \br'|$.  As the wedge geometry already lacks global translation and rotation symmetries, for universal properties it should be harmless to sacrifice these in the form of the potential.  Furthermore, $V(\br, \br')$ has the same scaling behavior as the Coulomb potential:
\begin{equation}
V(s \br, s \br') = \frac{1}{s} V(\br, \br') .
\label{eqn:new-potential-scaling}
\end{equation}
This can be shown by working with the definition Eq.~(\ref{eqn:new-potential}) and observing that
\begin{equation}
 \chi_{\lambda, n}(s \br) = \frac{1}{\sqrt{s}} \chi_{s\lambda, n}(\br) \text{.}
\end{equation}
Upon changing variables Eq.~(\ref{eqn:new-potential-scaling}) follows immediately.

\subsection{Superconducting ground plane}
\label{sec:ground-plane}

\begin{figure}
\includegraphics[width=8cm]{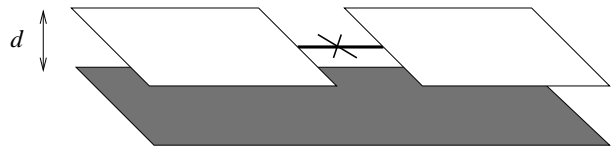}
\caption{Illustration of the system of superconducting films (white regions) and point contact, separated by an insulating layer of thickness $d$ from a superconducting ground plane (shaded region).  For the geometry shown, $\theta = \pi$.}
\label{fig:ground-plane}
\end{figure}

Here we consider a system of films and point contact in the presence of a superconducting ground plane.  As shown in Fig.~\ref{fig:ground-plane}, the ground plane is separated from the films by an insulating layer of thickness $d$.  The motivation to consider this setup is that the Coulomb interaction is screened with the ground plane in place, which allows $\alpha = 1$ to be realized in a system of superconducting (as opposed to superfluid) films.  Furthermore, this setup is quite favorable for the design of experiments that can be compared to our theory over a wide range of temperatures (see Sec.~\ref{sec:experiments}).

We model the ground plane as a superconducting film having stiffness
$\sg \gg \s$, as is appropriate for a thick layer of superconducting material.  Furthermore, we assume that the insulating barrier is sufficiently high and thick that Josephson tunneling between the films and the ground plane can be neglected.  For simplicity, we set the dielectric constant of the insulating medium to unity (\emph{i.e.}~vacuum).  The phase and number fields of the ground plane are denoted by $\varphi_g(\br)$ and $n_g(\br)$, respectively.

The goal is to derive an action like Eq.~(\ref{eqn:gaussian-phi-action}) for the phase difference across the contact.  It will turn out to be good enough to consider a problem in a simpler geometry, where we have only a single translation-invariant film and the ground plane (\emph{i.e.}~no contact).  In the limit $\sg \gg \s$, we conclude that the geometry of Fig.~\ref{fig:ground-plane} is well described by a model having no ground plane, and the short-range interaction of Eq.~(\ref{eqn:short-range-ints}), with superfluid velocity $v_s = \sqrt{16 \pi e^2 \s d}$.

%It will be possible to infer the form of the effective action from the phase correlations in the film.
%We will calculate the dispersion of the plasmon modes and show that there is one acoustic plasmon, %and one with $\omega \propto \sqrt{|\bk|}$ dispersion.  In the limit $\sg \gg \s$, the acoustic plasmon %propagates entirely in the film, while the other mode propagates only in the ground plane and does %not contribute to the transport.  Not surprisingly, the phase correlations in the film are identical to the %case of a film with short-range interactions.  Furthermore, because only the mode in the film enters, %the $\theta$-dependence of the phase correlations, and hence the effective action for the phase %difference across the contact, is also expected to be identical to the result for short-range interactions %and no ground plane.

The Hamiltonian for the coupled system of a single film and the ground plane is ${\cal H}_{{\rm bilayer}} = {\cal H}_K + {\cal H}_{{\rm int}}$, where
\begin{equation}
{\cal H}_K = \frac{\s}{2} \int d^2\br\, (\nabla \varphi)^2 + \frac{\sg}{2} \int d^2 \br\, (\nabla \varphi_g)^2 \text{,}
\end{equation}
and
\begin{equation}
{\cal H}_{{\rm int}} = \frac{(2 e)^2}{2} \int d^3\bR \, d^3\bR' \, \frac{N(\bR) N(\bR')}{|\bR - \bR'|} \text{.}
\end{equation}
Here we have defined the three-dimensional charge density
\begin{equation}
N(\bR) \equiv \delta(z - d/2) n(\br) + \delta(z + d/2) n_g(\br) \text{,}
\end{equation}
where $\bR = (\br, z)$ is the 3d position.  In momentum space,
\begin{equation}
{\cal H}_{{\rm int}} = \int \frac{d^3\bK}{(2\pi)^3} \frac{8 \pi e^2}{\bK^2} (n(\bk), n_g(\bk))
\left( \begin{array}{cc} 1 & e^{-i k_z d} \\
e^{i k_z d} & 1 
\end{array} \right)
\left( \begin{array}{c}
n(-\bk) \\
n_g(-\bk)
\end{array} \right) \text{,}
\end{equation}
where $\bK = (\bk, k_z)$.  The $k_z$-integral can be done via contour integration, yielding
\begin{equation}
{\cal H}_{{\rm int}} = \int \frac{d^2 \bk}{(2 \pi)^2} \frac{4 \pi e^2}{|\bk|} (n(\bk), n_g(\bk))
\left( \begin{array}{cc}
1 & e^{- d |\bk|} \\
e^{- d |\bk|} & 1
\end{array} \right)
\left( \begin{array}{c}
n(-\bk) \\
n_g(-\bk) 
\end{array} \right) \text{.}
\end{equation}
${\cal H}_{{\rm int}}$ simplifies in the basis $n_{\pm} \equiv (1/\sqrt{2})(n \pm n_g)$, where one obtains
\begin{equation}
{\cal H}_{{\rm int}} = \int \frac{d^2 \bk}{(2\pi)^2} \frac{4 \pi e^2}{|\bk|}
\Big[ (1 + e^{-d |\bk|})|n_+ (\bk)|^2 + (1 - e^{-d|\bk|}) |n_- (\bk)|^2 \Big] \text{.}
\end{equation}

At this point, it is straightforward to reformulate the problem in terms of an imaginary-time functional integral, integrating out $n_{\pm}(\bk)$ to obtain the following effective action for the phase fields:
\begin{equation}
S_{{\rm eff}} [\Phi] = \frac{1}{2\beta} \sum_{\on} \int \frac{d^2 \bk}{(2\pi)^2}
\Phi^T (\bk, \on) M_{{\rm eff}}(\bk, \on) \Phi(-\bk, -\on) \text{,}
\end{equation}
where $\Phi^T(\bk, \on) \equiv (\varphi (\bk, \on), \varphi_g (\bk, \on) )$ and
\begin{equation}
M_{{\rm eff}}(\bk, \on) = \left( \begin{array}{cc}
\frac{\on^2 |\bk|}{8 \pi e^2 (1 - e^{-2 d |\bk|})} + \s \bk^2 &
- \frac{\on^2 |\bk|}{16 \pi e^2 \operatorname{sinh}(d |\bk|)} \\
- \frac{\on^2 |\bk|}{16 \pi e^2 \operatorname{sinh}(d |\bk|)} &
\frac{\on^2 |\bk|}{8 \pi e^2 (1 - e^{-2 d |\bk|})} + \sg \bk^2
\end{array} \right) \text{.}
\end{equation}

The dispersion of the plasmon modes is given by solving $\operatorname{det} M_{{\rm eff}} (\bk, i \omega) = 0$.  This results in a quadratic equation for $\omega^2$, which yields the following two solutions when $d|\bk| \ll 1$:
\begin{eqnarray}
\omega(\bk) &=& v_s |\bk| \text{,} \\
\omega_g(\bk) &=& \sqrt{32 \pi e^2 (\s + \sg) |\bk|} \text{.}
\end{eqnarray}
The first of these is a linearly dispersing acoustic mode with velocity
\begin{equation}
v_s = \sqrt{\frac{\sg}{\s + \sg}} \sqrt{16 \pi e^2 \s d} \text{.}
\end{equation}
An analysis of the zero eigenvectors of 
$M_{{\rm eff}}(\bk, i \omega(\bk))$ shows that this mode propagates entirely in the film in the limit $\s \ll \sg$.  Similarly, the other plasmon mode propagates entirely in the ground plane in this limit.  In this limit
$v_s = \sqrt{16 \pi e^2 \s d}$. 

Therefore, when $\s \ll \sg$, the phase correlations in the film will be controlled entirely by the acoustic mode, and a treatment starting from a model with short-range interactions should give the correct action for the phase difference across the contact in the geometry of Fig.~\ref{fig:ground-plane}.  This expectation is substantiated by an explicit calculation, in the simpler translation-invariant geometry discussed above, of the correlation function 
\begin{equation}
C(\on) = \int_0^\beta d\tau \, e^{i \on \tau} \langle \varphi(\br = \boldsymbol{0}, \tau) \varphi(\br = \boldsymbol{0},0) \rangle \text{.}
\end{equation}
In general, this correlator for the phase at the point contact completely determines the desired effective action.  For small Matsubara frequency, the result is
\begin{equation}
C(\on) \sim \frac{1}{4\pi} \Big( \frac{1}{\s} + \frac{1}{\sg + \s} \Big) \ln \Big( \frac{\omega_c^2}{\on^2} \Big)
\text{,}
\end{equation}
which reduces to the result for a single film with short-range interactions when $\sg \gg \s$.

\section{Poor Contact}
\label{sec:weak-coupling}
We review the treatment in the poor-contact limit (\emph{i.e.}~$\tc \ll \s$), keeping in mind that such an approach should be viewed with caution.  We can use a Kubo formula to calculate ${\cal G}(T)$, the linear-response d.c.~conductance across the contact, to lowest order in $\tc$.  We shall show, below, that the resulting conductance appears to diverge below a characteristic temperature $T^*$.  These results agree with those of Ref.~\onlinecite{ybkim93}.  This calculation alone is not enough to say whether the divergence represents a true zero-resistance state, or simply a breakdown of perturbation theory in $\tc$.  The considerations in Sec.~\ref{sec:strongtcp} below demonstrate that the conductance is \emph{finite} for all $T > 0$, and, therefore, that weak-coupling perturbation theory fails below $T^*$.

%With the convention $e < 0$, the operator for the current flowing across the contact from plane 2 to plane 1 is:
The operator for the current flowing across the contact is
\begin{equation}
\hat{I} = 2 e \tc \sin (\hat{\phi}).
\end{equation}
Standard techniques of linear-response theory can be used to derive the Kubo formula relating the conductance to the retarded Green's function of the current:
\begin{equation}
{\cal G}(T, \omega) = - \operatorname{Re} \Big[ \frac{i}{\omega} G^I_R(\omega) \Big] \text{,}
\end{equation}
where
\begin{equation}
G^I_R(\omega) = 
-i \int_{-\infty}^{\infty} dt \Theta(t) e^{-i \omega t} 
\langle [ \hat{I}(t), \hat{I}(0) ] \rangle_{\tc = 0} \text{.}
\end{equation}
We shall be primarily interested in the d.c.~($\omega \to 0$) limit.

Straightforward manipulations lead to the expression
\begin{equation}
G^I_R(\omega) =
-i (2 e^2 \tc^2) \int_0^{\infty} dt e^{-i \omega t} 
\Bigg[ \exp\Big( - \langle \hat{\phi}^2 - \hat{\phi}(t)\hat{\phi}(0) \rangle \Big)
- (t \to -t) \Bigg] \text{.}
\end{equation}
Here, we have dropped terms involving $\exp(- \langle \hat{\phi}^2 + \hat{\phi}(t)\hat{\phi}(0) \rangle)$,
which vanish in the thermodynamic limit.  Inserting the result for the phase correlation function, obtained by analytic continuation in Appendix~\ref{app:phase-corr-fn}, we find
\begin{equation}
G^I_R(\omega) =
-4 e^2 \tc^2 \int_0^{\infty} dt e^{-i \omega t} \sin \big( B(t) \big) \exp \big( A(t) \big) \text{,}
\end{equation}
where
\begin{eqnarray}
A(t) &\equiv& \frac{\pi^2}{4 c \s} \int_0^{\omega_c} dx \big( \cos(x t) - 1 \big) \coth \Big( \frac{\beta x}{2} \Big) \text{,} \\
B(t) &\equiv& \frac{\pi^2}{4 c \s} \int_0^{\omega_c} dx \sin (x t) 
= \frac{\alpha}{\theta\s t} \label{eqn:b-of-t}\text{.}
\end{eqnarray}
In the second equality of Eq.~(\ref{eqn:b-of-t}) we have dropped a term oscillating at the cutoff frequency $\omega_c$, which is an unphysical artifact of the hard cutoff.  We now have, in the d.c.~limit,
%\begin{equation}
%{\cal G}(\omega, T) = \frac{4 e^2 \tc^2}{\omega} \int_0^{\infty} dt \sin (\omega t)
%\sin \Big( \frac{\alpha}{\theta\s t} \Big) \exp\big( A(t)\big) \text{,}
%\end{equation}
%or, in the d.c. limit
\begin{equation}
{\cal G}(T) = 4 e^2 \tc^2 \int_0^{\infty} dt\, t \sin \Big( \frac{\pi^2}{4 c \s t} \Big) \exp\big( A(t)\big) \text{.}\label{eqn:conductance}
\end{equation}

The integral of Eq.~(\ref{eqn:conductance}) can only diverge due to the contribution from long times.  To extract this behavior we first need to understand the behavior of $A(t)$ as $t \to \infty$.  In this limit, we have
$1/t \ll T \ll \omega_c$, and the dominant contribution to $A(t)$ comes from the region of the $x$-integral where  $1/t \ll x \ll T$.  There, the cosine is oscillatory and can be dropped, and $\coth(\beta x/2) \approx 2/\beta x$.  We then find that
\begin{equation}
%A(t) \sim -\frac{2 \alpha T}{\theta \s} \ln (t T) .
A(t) \sim - \frac{\pi^2 T}{2 c \s} \ln(t T) \text{.}
\end{equation}
At long times, the sine in Eq.~(\ref{eqn:conductance}) can be expanded to leading order, and the long-time contribution is thus
\begin{equation}
{\cal G}(T) \sim \frac{ \pi^2 e^2 \tc^2}{c \s} \int_{t_0}^{\infty} dt\,
(t T)^{- \pi^2 T/ 2 c \s} \text{,}
\end{equation}
where $t_0$ is arbitrary.  It is clear from this expression that ${\cal G}(T)$ apparently diverges for
\begin{equation}
T < T^* \equiv \frac{2 c \s}{\pi^2} \text{.}
%= \frac{\theta \s}{2 \alpha} .
\end{equation}

We shall show, below, that this divergence is due to a breakdown of perturbation theory, and the conductance is in fact always {\em finite} for $T > 0$.

\section{Good Contact}
\label{sec:strongtcp}

Given the apparent divergence of the conductance below $T^*$ in the limit of a poor contact, it will be fruitful to consider the opposite case of a good contact.  Even without knowing the poor-contact result, we should expect an expansion about the good contact limit to yield correct results, as discussed in the Introduction and Sec.~\ref{sec:formulation}.  Moreover, a variational calculation discussed in Appendix~\ref{app:variational} provides further support that the good-contact limit is always the correct starting point at low temperature.  In this Section, we first implement a duality transformation that allows us to work directly with the phase-slip degrees of freedom.  Following that, we
use the dual formulation to calculate the resistance of the point contact.

\subsection{Duality transformation}
\label{sec:duality}

%Even without knowing the poor-contact result, we should expect an
%expansion about the good contact limit to yield correct results, as
%discussed in the Introduction and Sec.~\ref{sec:formulation}.  

%Several routes to the dual theory are possible; here we describe the simplest, while
% in Appendix~\ref{app:lattice-duality} we carry out an alternate approach where a duality %transformation is made for the entire system \emph{before} integrating out the modes in the films.  %These approaches yield identical results, which gives us further confidence in their validity.

The starting point for the duality transformation is the partition function for the phase difference $\phi$ across the contact, 
$Z = \int {\cal D}\phi \exp(-S_0 - S_J)$, where $S_0$ and $S_J$ are defined in Eqs.~(\ref{eqn:gaussian-phi-action}) and (\ref{eqn:josephson-coupling}).
We proceed by replacing the cosine of $S_J$ with a Villain function:
\begin{equation}
\exp ( \tc \cos \phi) \to \sum_{\eta = -\infty}^{\infty} \exp\Big( - \frac{\bar{\tc}}{2} (\phi - 2\pi \eta)^2 \Big).
\end{equation}
This is merely a different $2\pi$-periodic potential, so the universal physics should be unaffected.  
To make this replacement in the partition function, we need only make $\eta$ a function of imaginary time (which we can take to satisfy periodic boundary conditions):
\begin{equation}
Z_{V} = \int {\cal D}\phi \sum_{\eta(\tau) = -\infty}^{\infty} \exp \Bigg[
 - \frac{2 c \s T}{\pi^2} \sum_{\on}|\fk(\on)|^2 \Big[\ln \Big( \frac{\on^2+\omega_c^2}{\on^2} \Big) \Big]^{-1}
 - \frac{\bar{\tc}}{2} \int_0^{\beta} d\tau \big(\phi(\tau) - 2\pi\eta(\tau) \big)^2 \Bigg] \text{.}
 \label{eqn:contact-villain-pf}
 \end{equation}
We have replaced the Josephson coupling $\tc$ by the energy $\bar{\tc}$.  The new partition function, with the cosine replaced by the Villain function, can be thought of as an effective description of the original one, and $\bar{\tc}$ as an {\em effective} $\tc$.  However, it is better to think of $\tc$ and $\bar{\tc}$ as parameters arising in two (slightly) different effective descriptions of the same underlying system.  Although we expect that $\tc \approx \bar{\tc}$, it is not important to have a more precise relationship between these parameters, or even to distinguish them, unless 
one of them can be directly measured.  Indeed, in the absence of such a measurement, $\bar{\tc}$ must be treated as a fitting parameter in the resistance formula of Eq.~(\ref{eqn:resistance}).

To specify $Z_V$ more precisely, it is useful to write $\eta(\tau)$ in terms of its time derivative $\rho(\tau)$, which is a sum of delta functions at those times when phase slips occur.  Because we take $\eta(\tau) = \eta(\tau + \beta)$, an equal number of ``positive'' ($\eta \to \eta + 1$) and ``negative'' ($\eta \to \eta - 1$) phase slips occur in each of the configurations summed over.  In a configuration having $n$ positive phase slips, we denote their times by $\tau_i$, and the times of their negative counterparts by $\bar{\tau}_i$.  Then we have:
\begin{equation}
\rho(\tau) = \sum_{i = 1}^{n} \Big ( \delta(\tau - \tau_i) - \delta(\tau - \bar{\tau}_i) \Big) .
\end{equation}
Without loss of generality, we can take $\eta(\tau = 0) = 0$ and write 
$\eta(\tau) = \int_0^\tau d\tau' \rho(\tau')$.  In the Villain partition function, the sum over $\eta$ becomes a sum over $\rho$, which is easy to define precisely:
\begin{equation}
\sum_{\rho(\tau)} \equiv \sum_{n=0}^{\infty} \Big(\frac{t_v^n}{2^n n!} \Big)^2
\int d\tau_1 \dots d\tau_n \int d\bar{\tau}_1 \dots d\bar{\tau}_n .
\end{equation}
Here, it was necessary to introduce an energy $t_v$, which is the fugacity per unit imaginary time for phase-slip events.  The factor of $2^{-2n}$ is inserted for later convenience.  Upon transforming to Fourier space and integrating out $\phi$, we obtain a partition function only in terms of $\rho$:
\begin{equation}
Z_V = \sum_{\rho(\tau)} \exp \Bigg[
-\frac{T}{2} \sum_{\on} \frac{\bar{\tc}}{1 + \frac{\pi^2 \bar{\tc}}{4 c \s}  \ln \Big( \frac{\on^2 + \omega_c^2}{\on^2} \Big)}  \Big| \frac{2\pi \tilde{\rho}(\on)}{\on} \Big|^2 \Bigg] \text{.}
\label{eqn:rho-action}
\end{equation}

The duality transformation is completed by observing that Eq.~(\ref{eqn:rho-action}) is equivalent to the following sine-Gordon theory:
\begin{equation}
Z_{{\rm dual}} = \int {\cal D}\theta \exp \Bigg\{
-\frac{T}{2} \sum_{\on} \Big[ \frac{1}{\bar{\tc}} + \frac{\pi^2}{4 c \s}  \ln \big( \frac{\on^2 + \omega_c^2}{\on^2} \big)
\Big] \on^2 |\tilde{\theta}(\on)|^2
+ t_v \int d\tau \cos(2\pi \theta) \Bigg\}
\label{eqn:dual-theory}
\end{equation}
The equivalence is simple to establish: expanding in powers of $t_v$ and denoting the
Gaussian part of the action by $S^{{\rm dual}}_0[\theta]$, we obtain the partition function
\begin{equation}
Z_{{\rm dual}} = \sum_{n=0}^\infty \frac{t_v^n}{2^n n!} \sum_{\sigma_1 \cdots \sigma_n = \pm 1}
\int d\tau_1 \cdots d\tau_n \int {\cal D}\theta
\exp \Big[ - S^{{\rm dual}}_0[\theta] + 2\pi i \sum_{i=1}^n \sigma_i \theta(\tau_i) \Big] \text{.}
\end{equation}
At this point it is helpful to note that $\theta(\tau) \to \theta(\tau) + c$ is a symmetry of $S^{{\rm dual}}_0$, which means that only ``neutral'' configurations, \emph{i.e.}~those having $\sum_i \sigma_i = 0$, contribute.  This allows us to reorganize the above expression and write:
\begin{equation}
Z_{{\rm dual}} = \sum_{\rho(\tau)} \int {\cal D}\theta \exp \Big[ -S^{{\rm dual}}_0[\theta] + 2\pi i \int d\tau \rho(\tau) \theta(\tau) \Big] \text{.}
\end{equation}
Upon integrating out $\theta$ we immediately recover Eq.~(\ref{eqn:rho-action}).

It is important to note that the term proportional to $1/\bar{\tc}$ in the dual action is dominated by the logarithm at low frequency.  Therefore, $\bar{\tc}$ will not not play a role in the low-temperature limit.  However, it is important to keep this term in the formulation.  This permits one to access the $\tc \ll \s$ and $T \gg T_J$ regime discussed in Sec.~\ref{sec:intro}, where the resistance displays the truly activated behavior (\emph{i.e.}~having no logarithmic temperature-dependence) of Eq.~(\ref{eqn:purely-activated-barrier}).

From these considerations, we see that the operator $\exp(2\pi i \theta(\tau))$ inserts a phase-slip event at time $\tau$.  The $t_v \cos(2\pi \theta)$ term in Eq.~(\ref{eqn:dual-theory}) can therefore be interpreted as a vortex-hopping process.   Furthermore, because phase and number are canonically conjugate variables, we can think of $\theta$ as a charge difference between the films; in fact, it is precisely the \emph{charge imbalance} $Q_{{\rm imb}}$ between the
two films (\emph{i.e.}~half the difference of the total charges of the films):
\begin{equation}
\theta = \frac{1}{4e}(Q_2 - Q_1) \equiv \frac{1}{2e} Q_{{\rm imb}} \text{,}
\end{equation}
where $Q_i$ is the total charge on film $i$.
Therefore, the operator for the voltage drop across the contact  (from film 1 to film 2) is given by:
\begin{equation}
V = - \frac{\delta}{\delta Q_{{\rm imb}}} \big( - t_v \cos (2\pi \theta) \big)
 =  \frac{1}{2e} \frac{\delta}{\delta \theta} \big( t_v \cos (2\pi \theta) \big)
= - \frac{\pi t_v}{e} \sin (2 \pi \theta) \text{.}
\end{equation}

\subsection{Calculation of the resistance}
\label{sec:resistance-calculation}

In order to find the resistance to lowest order in $t_v$, we need to calculate the response of the voltage to a weak external {\em current}, using the dual partition function Eq.~(\ref{eqn:dual-theory}).  This can be done by working in real time and replacing $\theta \to \theta - I t/2 e$ in the vortex-hopping term [\emph{i.e.} the $t_v \cos(2\pi\theta)$ term].  Here, $I$ is the externally-controlled current flowing across the contact from film 1 to film 2.  The situation is now completely analogous to that of the standard derivation of the Kubo formula for a conductance, and we obtain for the resistance:
\begin{equation}
R(\omega, T) = - \operatorname{Re} \Big[ \frac{i}{\omega} G^V_R(\omega) \Big].
\end{equation}
Here, $G^V_R(\omega)$ is the real-time retarded Green's function for the voltage:
\begin{eqnarray}
G^V_R(\omega) &=& -i \int_{-\infty}^{\infty} dt \Theta(t) e^{-i \omega t} 
\langle [ \hat{V}(t), \hat{V}(0) ] \rangle_{t_v = 0} \\
&=& -i \frac{\pi^2 t_v^2}{2 e^2} \int_0^{\infty} dt e^{-i \omega t}
\big( e^{F(t)} - e^{F(-t)} \big) \text{,}
\end{eqnarray}
where
\begin{eqnarray}
F(t) &\equiv&  4 \pi^2 \langle \hat{\theta}(t) \hat{\theta}(0)-  \hat{\theta}^2 \rangle \\
&=& 16 c \s \int_0^{\omega_c} \frac{d x}{x^2} 
\Bigg( \big(\cos(x t) - 1 \big) \coth \Big( \frac{\beta x}{2} \Big) - i \sin (x t) \Bigg)
\frac{1}{\pi^2 + \Big[ 4 c \s / \pi^2 \bar{\tc} + \ln \big( (\omega_c / x)^2 - 1\big) \Big]^2} .
\end{eqnarray}
This correlator is obtained by the analytic continuation of its imaginary-time analog; see Appendix~\ref{app:imag-time-dual-cf}.  
After straightforward manipulations, the d.c.~resistance can be written as
\begin{equation}
R(T) = \frac{i \pi^2 t_v^2}{2 e^2} \int_{-\infty}^{\infty} dt \, t \, e^{F(t)} \text{.}
\label{eqn:dc-resistance}
\end{equation}
We note that $F(t)$ is analytic in the entire complex plane.

Let us first consider the case of zero temperature, where we can write
\begin{equation}
F(t)  = 16 c \s \int_0^{\omega_c} \frac{d x}{x^2} (e^{-i x t} - 1)
\frac{1}{\pi^2 + \Big[ 4 c \s / \pi^2 \bar{\tc} + \ln \big((\omega_c / x)^2 - 1\big) \Big]^2} \text{.}
\end{equation}
We shall continue the resistance integral, Eq.~(\ref{eqn:dc-resistance}), into the lower half-plane.  If we let $t = |t| e^{i \gamma}$, where
$\gamma \in [-\pi,0]$, it is easy to show that, up to logarithmic corrections,
\begin{equation}
\operatorname{Re} \big[ F(t) \big] \sim - a(\gamma) |t| \text{,}
\end{equation}
for $|t| \to \infty$, where $a(\gamma) > 0$.  Therefore, the integrand of Eq.~(\ref{eqn:dc-resistance}) decays exponentially in this limit, and, by deforming the $t$-contour arbitrarily far into the lower half-plane, we see that $R(T = 0) = 0$.

Next, we consider the case of nonzero temperature, for which it is convenient to write
\begin{equation}
F(t) = 8 c \s \int_0^{\omega_c} \frac{dx}{x^2} \frac{1}{\sinh \big(\frac{\beta x}{2}\big)}
\Big[e^{i x(t + i\beta/2)} + e^{-i x(t + i\beta/2)} - 2 \cosh \big(\frac{\beta x}{2}\big) \Big]
\frac{1}{\pi^2 + \big[4 c \s / \pi^2 \bar{\tc} + \ln \big((\omega_c / x)^2 - 1\big) \big]^2} \text{.}
\end{equation}
Again we allow $t$ to be complex and consider $F(t_r + i t_i)$ for $t_i \in [-\beta/2,0]$, and again up to logarithmic corrections we find that $\operatorname{Re}[F(t)]$ goes linearly to $-\infty$ for $t_r \to \pm \infty$.  Then, instead of integrating along the real axis, we can integrate along the contour swept out by $t  - i\beta/2$ for real $t$ to obtain
\begin{equation}
\label{eqn:continued-dc-resistance}
R(T) = \frac{\pi^2 t_v^2}{4 e^2 T} \int_{-\infty}^{\infty} dt \, e^{f(t)} \text{,}
\end{equation}
where
\begin{equation}
f(t) = F(t - i \beta/2) = 16 c \s \int_0^{\omega_c} \frac{dx}{x^2}
\frac{\cos(x t) - \cosh \big( \frac{\beta x}{2} \big) }{\sinh \big(\frac{\beta x}{2}\big)}
\frac{1}{\pi^2 + \big[ 4 c \s / \pi^2 \bar{\tc} + \ln \big((\omega_c/x)^2 - 1\big) \big]^2} \text{.}
\label{eqn:defn-of-f-function}
\end{equation}

At low temperature, the resistance integral (\ref{eqn:continued-dc-resistance}) can be evaluated by the saddle-point method.  It is easy to see that $f(0) \to -\infty$ in the $T \to 0$ limit and, furthermore, that
$f(0) > f(t)$ for all $t$ (and $T$) -- this means that $t=0$ is the global maximum of the integrand.  We then have 
\begin{equation}
R(T) \approx \frac{\pi^2 t_v^2}{4 e^2 T}e^{f(0)} \int_{-\infty}^\infty \exp\big(\frac{1}{2} f''(0) t^2 \big) =
 \frac{\pi^2 t_v^2}{4 e^2 T} \sqrt{\frac{2\pi}{-f''(0)}} \exp\big( f(0) \big)\text{,}
\label{eqn:saddle-point-resistance} 
\end{equation}
which is expected to be asymptotically exact as $T \to 0$.  
As shown in Appendix~\ref{app:integrals-saddle-pt-validity}, a conservative criterion for the validity of the saddle-point evaluation is that $f(0)$ should be large and negative.  This means that we should expect Eq.~(\ref{eqn:saddle-point-resistance}) to be a good approximation for the resistance whenever $f(0) \lesssim -1$.
 
We then need to evaluate the low-temperature form of $f(0)$ and $f''(0)$ -- this is done in Appendix~\ref{app:integrals}.  The results are:
\begin{equation}
 f(0) \approx - \frac{ c \s}{T} \frac{1}{\ln (\omega_c / 2 T) + 2 c \s / \pi^2 \bar{\tc} }  = - \frac{E_A(T)}{T}
 \text{,}
 \label{eqn:f0-formula}
 \end{equation}
 and
 \begin{equation}
 f''(0) \approx - 8 c \s T \frac{1}{\ln (\omega_c / 2 T ) + 2 c \s / \pi^2 \bar{\tc} } = -8 T E_A(T)
  \text{.}
 \label{eqn:f0pp-formula}
 \end{equation}
These relations are asymptotically exact as $T \to 0$, where the dependence on $\bar{\tc}$ drops out.  
More precisely, the \emph{ratio} of Eq.~(\ref{eqn:f0-formula}) to the exact value of $f(0)$ goes to unity as $T \to 0$.  However, the \emph{difference} of these two quantities does not go to zero; this is due to additive corrections to $f(0)$ that are proportional to $1/(T \ln^2(T))$, which are sub-dominant but nonetheless diverge at zero temperature.  These statements also hold for $f''(0)$ and Eq.~(\ref{eqn:f0pp-formula}).

We show in Appendix~\ref{app:integrals-poor-contact} that Eqs.~(\ref{eqn:f0-formula}) and~(\ref{eqn:f0pp-formula}) are also asymptotically exact in the extreme good and poor contact limits.  Precisely, in the good-contact limit we require that $\bar{\tc} \gg \s$ and $T \ll \s$.  In this case, we recover the universal low-temperature forms for $f(0)$ and $f''(0)$, but for all $T \ll \s$.  In the poor-contact limit we require that $\bar{\tc} \ll \s$ and $T_J \ll T \ll \bar{\tc}$. 
We then have
\begin{equation}
f(0) \sim - \frac{\pi^2 \bar{\tc}}{2 T} \text{,}
\label{eqn:poor-contact-f0-formula}
\end{equation}
and
\begin{equation}
f''(0) \sim -4 \pi^2 T \bar{\tc} \text{.}
\end{equation}
Note that we must make the restriction $T \gg T_J$, because when $T \lesssim T_J$ the resistance crosses over to the universal low-temperature result.  

The expressions Eq.~(\ref{eqn:f0-formula}) and Eq.~(\ref{eqn:f0pp-formula}) can thus be thought of as an interpolation between two limits where they are exact.
Although this interpolation between good- and poor-contact limits is approximate, these formulas are valid to reasonable accuracy over a wide range of parameters.  For example, using the parameters of the setup (with ground plane) discussed in Sec.~\ref{sec:experiments}, a straightforward numerical evaluation of $f(0)$ shows that the formula Eq.~(\ref{eqn:f0-formula}) agrees to within $15 \%$ for $T \leq 5\,{\rm K}$.  However, if \emph{either} the universal low-temperature form ($\bar{\tc} / \s \to \infty$) or the poor-contact form Eq.~(\ref{eqn:poor-contact-f0-formula}) is used, there is only about $50 \%$ agreement at $T \approx 5\, {\rm K}$.

\section{Mechanical analogy}
\label{sec:mechanical}

The resistance formula Eq.~(\ref{eqn:resistance}) can be simply interpreted in terms of an analogy to an essentially classical system of springs at finite temperature.  We focus here only on the exponential
part of the resistance, which gives the dominant temperature dependence.  We write this in the form
\begin{equation}
R(T) \propto \exp \Big( -\frac{(1/T)}{(1/ c \s)\ln(\omega_c / 2 T) + 2 / \pi^2 \bar{\tc}} \Big) \text{.}
\label{eqn:resistance-activated-form}
\end{equation}

In the limit of large Josephson coupling $\bar{\tc}$ (or at very low temperature),  the resistance becomes 
\begin{equation}
R(T) \propto \exp \Big( -\frac{c \s}{T \ln(\omega_c / 2 T)} \Big) \text{.}
\end{equation}
Except for the logarithm, the resistance is activated in temperature.
On the other hand,  when $\bar{\tc}$ is small and the temperature is not too low, we have
\begin{equation}
R(T) \propto \exp \Big( - \frac{\pi^2 \bar{\tc}}{2 T} \Big) \text{,}
\end{equation}
which really is an activated form.  

The above crossover between two kinds of activated behavior suggests that we might be able to understand the resistance in terms of the thermally activated dynamics of an effective classical system.  We shall obtain a description of such a system starting from the action for the phases $\Phi_i \equiv \varphi_i(0)$ at the contact -- it will be instructive to retain the phases separately here, rather than focusing only on the phase difference $\phi$.  The action takes the form $S = S^1_0 + S^2_0 + S_J$, where
\begin{equation}
S^i_0 = \frac{4 c \s T}{\pi^2} \sum_{\omega_n} \left[ \ln \frac{\omega_n^2 + \omega_c^2}{\omega_n^2} \right]^{-1} | \tilde{\Phi}_i(\omega_n)|^2 \text{.}
\end{equation}
We wish to replace the slowly-varying logarithm by an $\on$-independent constant.  As $\on \in 2\pi T {\mathbb Z}$, we have, very roughly,
\begin{equation}
\ln \frac{\on^2 + \omega_c^2}{\on^2} \approx 2 \ln \frac{\omega_c}{T} \text{.}
\end{equation}
The term $S_J$ is the Josephson coupling, which we expand to quadratic order:
\begin{equation}
S_J \approx \frac{\bar{\tc}}{2} \int_0^\beta d\tau \big( \Phi_1(\tau) - \Phi_2(\tau) \big)^2 \text{.}
\end{equation}
Furthermore, we imagine that the phase deep within each film is fixed to a classical value $\bar{\Phi}_i$.

\begin{figure}
\includegraphics[width=8cm]{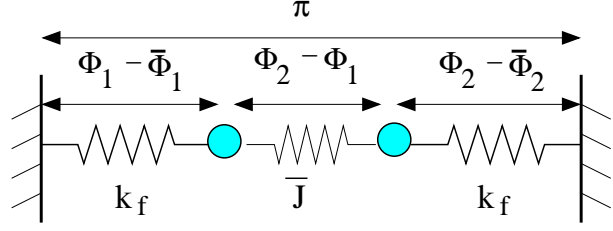}
\caption{Depiction of the classical spring system defined by the action~(\ref{eqn:spring-action}).  
The spring constants of the left and right springs are determined by the power-law phase fluctuations in the films [$k_f = 4 c \s / \pi^2 \ln(\omega_c / T)$]. The central spring represents the Josephson coupling
and has spring constant $\bar{\tc}$.
For a phase slip to occur, the system must cross over the energy barrier associated with stretching the coupled system a distance $\pi$.
\label{fig:springs}}
\end{figure}

With these approximations, we arrive at the action
\begin{equation}
S \approx \int_0^\beta d\tau \Big[  
\frac{2 c \s}{\pi^2 \ln(\omega_c / T)} \big( (\Phi_1 - \bar{\Phi}_1)^2 + (\Phi_2 - \bar{\Phi}_2)^2 \big)
+ \frac{\bar{\tc}}{2} \big( \Phi_1 - \Phi_2 \big)^2 \Big] \text{.}
\label{eqn:spring-action}
\end{equation}
Note that we do not explicitly account for the $2\pi$ periodicity; more precisely, if we had been more careful, the action would be independent of $n$ if $\bar{\Phi}_2 = \bar{\Phi}_1 + 2\pi n$.
This action describes a simple mechanical picture of three springs coupled in series, as shown in Fig.~\ref{fig:springs}.  In order for a phase slip to occur across the contact, the system must tunnel over an energy barrier given by setting $\bar{\Phi}_2 - \bar{\Phi}_1 = \pi$.  This is equivalent to stretching the combined system of three springs by a distance $\pi$, for which the excitation energy is found to be
\begin{equation}
E = \frac{1}{2}k_{{\rm eff}}(\pi)^2 
= \frac{1}{\frac{2}{\pi^2 \bar{\tc}} + \frac{1}{c \s} \ln \Big(\frac{\omega_c}{T} \Big) } \text{,}
\label{eqn:effective-spring-const}
\end{equation}
where we have used the rule for adding spring constants in series: 
\begin{equation}
k^{-1}_{{\rm eff}}= \sum_i k^{-1}_i \text{.}
\end{equation}

In this picture, the resistance is proportional to the Boltzmann weight of the activation barrier
$\exp(-E/T)$, so we recover Eq.~(\ref{eqn:resistance-activated-form}).  We thus see that it is possible, at least roughly, to think of the low-temperature resistance as arising due to the thermal activation over an effective, temperature-dependent barrier.  The properties of the barrier are determined by the long-wavelength physics of the superfluid films (and also their geometry, via the opening angle).  That one does not truly have simple activated dynamics is evident in the
logarithmic temperature-dependence of the barrier height, which arises from the QLRO in the films.

\section{Relation to experiments}
\label{sec:experiments}

In this section we discuss the relevance of our theory to experiments on superconducting films.  Although we consider only systems in the wedge geometry discussed explicitly in the paper, we expect that the basic  features of our results will apply to a much broader class of Josephson tunneling experiments into and between 2d superconductors.  Indeed, we hope to stimulate more experimental and theoretical interest in such systems; some speculations on future directions such work might take are contained in Sec.~\ref{sec:discussion}.

The most striking feature of the resistance formula [Eq.~(\ref{eqn:resistance})] is the effective activation barrier $E_A(T)$.  In the universal limit, \emph{i.e.}~$\s/J \ll 1$ or $T \to 0$, this is set primarily by the superfluid stiffness in the films.  The stiffness can be independently determined in thin film samples, via an inductance measurement,\cite{fiory83} so it should be possible to test this prediction rather thoroughly.  Furthermore, $\s$ could potentially be varied \emph{in situ\/} by tuning an in-plane magnetic field.
The barrier also depends on $\omega_c$ and $\bar{\tc}$, which should really be considered fitting parameters.
We can estimate $\omega_c$ via $\omega_c \approx \omega(|\bk| = 2\pi/\xi)$, where $\xi$ is the coherence length.  
%For the setup with a superconducting ground plane, we should replace $\xi$ with the film-plane %separation $d$ (as long as $d > \xi$).
In many cases, $\bar{\tc}$ can also be estimated; for example, consider a junction that is a constriction of length $\ell$ and width $w$ between two films each having bulk superfluid stiffness $\s$.  Then we estimate $\bar{\tc} \approx w \s/ \ell$.  These estimates constrain the range of values of $\omega_c$ and $\bar{\tc}$ that may reasonably be used in fitting to experimental data.

The most interesting regime is the universal limit [Eq.~(\ref{eqn:universal-lowT-barrier})], where the dependence on $\tc$, and hence on the details of the contact, disappears.  In general, it may be difficult to access this regime by lowering the temperature;  one needs $\ln (\hbar \omega_c / 2 k_B T) \gg 4 c \s / \pi^2 \bar{\tc}$, and the logarithm increases only slowly as $T$ is decreased.
Instead, it would probably be easier to fabricate contacts having large $\tc / \s$; this could be achieved, for example, by fabricating short and wide constrictions between thin films.  It would also be interesting to look for the purely activated behavior that should be present over a wide temperature-range when $\tc / \s$ is small.

Another important consideration in fitting to Eq.~(\ref{eqn:resistance}) is that it may be better to use the more accurate form that can be obtained by numerically evaluating the appropriate integrals, especially when $\tc \approx \s$.  As discussed in Sec.~\ref{sec:resistance-calculation}, 
Eq.~(\ref{eqn:resistance}) is obtained from
\begin{equation}
R(T)  =
 \frac{\pi^2 t_v^2}{4 e^2 T} \sqrt{\frac{2\pi}{-f''(0)}} \exp\big( f(0) \big)\text{,}
\end{equation}
where
\begin{equation}
f(t) = \frac{8 c \s}{T} \int_0^{\beta \omega_c / 2} \frac{du}{u^2}
\frac{\cos(2 u t / \beta) - \cosh(u) }{\sinh(u)}
\frac{1}{\pi^2 + \big[ 4 c \s / \pi^2 \bar{\tc} + \ln \big((\beta\omega_c / 2 u)^2 - 1\big) \big]^2} \text{.}
%f(t) =  4 \pi^2 E_A \int_0^{\omega_c} \frac{dx}{x^2}
%\frac{\cos(x t) - \cosh \big( \frac{\beta x}{2} \big) }{\sinh \big(\frac{\beta x}{2}\big)}
%\frac{1}{\pi^2 + \big[ E_A / \bar{\tc} + \ln \big((\omega_c/x)^2 - 1\big) \big]^2} \text{;}
\end{equation}
It is important to work with this more accurate form because the corrections to $f(0)$ not contained in Eq.~(\ref{eqn:resistance}) are proportional to $1/T \ln^2(T)$; these are dominated only rather weakly by the $1/T \ln(T)$ term, and they also diverge at low temperature.  We note, however, that the form~(\ref{eqn:resistance}) is already rather accurate over a wide range of temperatures.  For example, using the parameters discussed below in the setup involving a ground plane, the approximate value of $f(0)$ is within $15\%$ of its exact value for all $T < 5 {\rm K}$.

A crucial issue is to determine whether the system is indeed in a regime where our theory applies.  To illustrate the important considerations, we consider a system that has already been fabricated and studied by Chu, Bollinger and Bezryadin,\cite{chu04} in which two MoGe thin films are connected by a narrow constriction.  The films have a characteristic linear dimension $L$ and exhibit a bulk BKT transition at $T_{{\rm BKT}} \approx 5\, {\rm K}$.  For concreteness, we assume that $\s/k_B \approx 10\,{\rm K}$ for $T \ll T_{{\rm BKT}}$, although in practice the stiffness should be determined by a direct measurement.
We have opening angle $\theta = \pi$ and Coulomb interactions (\emph{i.e.}~$\alpha = 2$).  The fabricated constrictions discussed in Ref.~\onlinecite{chu04} have width $w \approx 20\,{\rm nm}$ and length $\ell \approx 100\, {\rm nm}$.  The coherence length is about $\xi \approx 7\,{\rm nm}$.  With these parameters, we have $\hbar \omega_c / k_B \approx 2000\,{\rm K}$ and $\bar{\tc} / k_B \approx 2\,{\rm K}$.

We also consider a modification of the above setup, in which a superconducting ground plane is  separated from the superconducting films by an insulating layer of thickness $d = 100\,{\rm nm}$.  As discussed in Sec.~\ref{sec:ground-plane}, the interactions are screened down to a short-range form 
and, with the other parameters as above, we have the superfluid velocity
$v_s = (1 / \hbar)\sqrt{16 \pi e^2 \s d} \approx 3.8 \times 10^8 \, {\rm cm}/{\rm s}$.  In this case, provided $d > \xi$, the cutoff is set by $d$ rather than $\xi$; that is, $\omega_c = \omega(k = 2\pi/d)$.  For our parameters, we have $\hbar \omega_c / k_B \approx  2000\, {\rm K}$.

In real systems, such as these, when does the resistance formula apply?  First, the resistance should be in some sense small for our result to be a good approximation.  More precisely, we require that $E_A(T) / k_B T \gtrsim 1$, so that terms of higher-order in $t_v$ are not important.  For both parameter sets discussed above, this is satisfied for $T \lesssim 5\, {\rm K}$.  

Second, the films must be in the thermodynamic limit.  That is, they must be large enough that we can safely neglect finite-size effects, as well as the influence of the outside environment.  This will be the case when plasmons undergo thermal decoherence within the films much faster than they can travel across it.  For films of characteristic linear dimension $L$, this is the case for $T \gg T_L$, where $T_L$ is a crossover temperature defined by
\begin{equation}
k_B T_L = \hbar \omega(k = 2\pi / L) \text{.}
\label{eqn:TL-definition}
\end{equation} 
It should be noted that $T_L$ decreases with increasing $L$.  For acoustic plasmons, the definition of $T_L$ simply comes from a comparison of the thermal decoherence time $t_T = 2\pi \hbar / k_B T$, and the time to travel across the system, $t_L = L / v_s$.  For plasmons having $\omega \propto \sqrt{k}$ dispersion, one has to replace $v_s$ by the group velocity of the fastest plasmons in the system, which is given by $v_g^{{\rm max}} = \frac{1}{\hbar} \frac{ d\omega}{d k}(k = 2\pi/L)$.  This actually results in Eq.~(\ref{eqn:TL-definition}) but with an extra factor of $1/2$ on the right-hand side.  However, one should also impose the condition that the lowest-energy plasmons have significant thermal occupation, which holds above $T_L$ as defined in Eq.~(\ref{eqn:TL-definition}).  In any case, it is more important simply to have an order of magnitude estimate for $T_L$.

A separate finite-size effect is the Coulomb blockade phenomenon, where the transport is dominated by the charging energy $E_c$ required to add a single Cooper pair to one of the films, and insulating behavior results.  This physics is only important when $T \lesssim E_c / k_B$.  Noting that $E_c \propto 1/L^2$ for short-range interactions, and that $E_c \propto 1/L$ for Coulomb interactions, in both cases one has $T_L \gg E_c / k_B$ for large enough films.  Therefore, it is enough to require that $T \gg T_L$ to eliminate finite-size effects, \emph{including} the Coulomb blockade.

With a superconducting ground plane as above, $T_L \approx 2 \times 10^{-2}\, {\rm K}$ for $L = 1\,{\rm cm}$.  On the other hand, without the ground plane, $T_L \approx 1\, {\rm K}$ for $L = 1\, {\rm cm}$.   There are two lessons to be learned from this.  The first is that the ground plane is quite helpful in lowering $T_L$.  The second is that, even in with the ground plane in place, it is important to have rather large films for our theory to apply over a wide range of temperatures.  Without the ground plane, it may be possible to achieve a modest reduction in $T_L$ by sandwiching the system, above and below, by an insulating medium having a rather large dielectric constant $\epsilon$, as $T_L \propto \epsilon^{-1/2}$.

\section{Discussion}
\label{sec:discussion}

In this paper we have derived a general formula for the resistance of a
point contact between two infinite superconducting films. This quantity 
is a measure of the strength and nature of fluctuations
in the superconductors.  The low-temperature resistance is very nearly activated, with an activation barrier $E_A$ determined only by the superfluid stiffness of the films and simple features of the geometry and interactions between Cooper pairs.  Deviations from purely activated behavior enter via a weakly temperature-dependent reduction of the barrier height that depends only logarithmically on temperature.  This result complements, and is intermediate between, the much-studied physics of 3d bulk electrodes, where the Josephson effect occurs ($R = 0$), and 1d superconducting wires, where Luttinger-liquid physics leads to a power-law tunneling resistance ($R \propto T^\alpha$).  In the 2d setting, the quasi-long-range order below $T_{{\rm BKT}}$ is not strong enough to enforce zero resistance at nonzero temperature.  However, phase fluctuations are strongly suppressed, and the nearly activated resistance is quite close to a true Josephson effect.

As discussed in detail in Sec.~\ref{sec:experiments}, it should be possible to test our results directly, via measurements of point-contact tunneling transport between superconducting thin films.  More generally, 
we expect that the basic features of our results -- most notably, the nearly-activated form of $R(T)$ -- will survive in a broader variety of systems involving tunneling into and between 2d superconductors and superfluids.  It would be interesting to understand how this physics plays out in geometries more complicated than the one considered here, perhaps involving multiple contacts, multiple films, and other ``circuit elements.''  

%It would be particularly intriguing if it is possible to control the barrier height in a single sample by %tuning an easily accessible parameter such as a magnetic field or a gate voltage.

In addition to enhancing our understanding of QLRO in usual
superconducting films, superconducting point-contact measurements could perhaps be used as a
probe of fluctuating off-diagonal order in poorly-understood systems, such as the
high-$T_c$ cuprates, and disordered films near a
superconductor-insulator transition. Our work
provides a baseline for such measurements, rooted in the well-understood physics of the superconducting state below $T_{{\rm BKT}}$.
Many interesting strongly-correlated and disordered materials possess gapless quasiparticle excitations -- for example, disorder and field-induced
quasi-particles in amorphous films or nodal
quasiparticles in the cuprates -- to which a point-contact measurement would be sensitive.
In strongly disordered systems, the geometry could change in a nontrivial way, 
for instance if the QLRO in the film is sustained on a superconducting cluster near the percolation threshold.  It would be very interesting to
carry out analogous tunneling resistance calculations for the above
cases.

Even in the simple system considered here, several interesting issues remain to be addressed.
One important piece of the physics not discussed here is the
zero-temperature \emph{nonlinear\/} I-V curve. This would give an additional
measure of the off-diagonal order.  It would also be useful to develop a better understanding how the phase-slip tunneling amplitude $t_v$, which plays a crucial role in our analysis, depends on the more fundamental parameters of the system.  Finally, we remark that a renormalization-group treatment has contributed substantially to the understand of the 1d analog of this problem.  So far, such an approach has not been successful here, but an analysis along these lines may be possible.

%In our calculation we introduced
%the phase slip fugacity, $t_V$. We understand qualitatively how $t_V$
%depentds on $\tc$, $\s$ and $T$, but in order to get full unerstanding
%of the point-contact problem, a quantitative understanding of $t_V$
%will also be quite useful.   

\begin{acknowledgments}
We are grateful to Leon Balents, Alexey Bezryadin, Arun Paramekanti and Xiao-Gang Wen for useful discussions.  This research is supported by the Department of Defense NDSEG program (M.H.); NSF
Grant No. PHY99-07949 (G.R. and M.P.A.F.); NSF Grant No. DMR-0210790 (M.P.A.F.); and the U.S. Department of Energy, Division of  Material Sciences under Award No.~DEFG02-91ER45439 [through the Frederick Seitz Materials Research Laboratory at UIUC] (P.M.G.).
\end{acknowledgments}

\appendix
\section{Effective action for the point contact}
\label{app:actions}

In this appendix we derive the effective action for the phase difference across the point contact for the various geometries and interactions of interest.  Most of these manipulations are completely straightforward; however, in the case of Coulomb interactions in the wedge geometry (Sec.~{\ref{sec:lr-wedge}), we have found it necessary to modify the form of interaction to proceed analytically.  As discussed in more detail in Sec.~\ref{sec:formulation}, we believe that our modification does not affect the universal physics or the form of Eq.~(\ref{eqn:gaussian-phi-action}).

%In Sec. (\ref{sec3}) we derived the linear response of the point
%contact in the weak and strong coupling regime using a real-time
%formulation of the problem. Here we show how these
%correlation functions are obtained by continuation of the imaginary
%time correlation functions. We also demonstrate how to obtain the
%action in terms of the dual variable, and the correlation functions
%involved. The imaginary time approach is particularly important for
%the variational approach in Sec. (\ref{variational}).

\subsection{Bilayer geometry, short-range interactions}

The imaginary-time action for the planes and contact is $S = S_1 + S_2 + S_J$, where
\begin{equation}
S_i = \intt_0^\beta d\tau \int d^2 \br \l( \frac{\s}{2} \l( \nabla \varphi_i \r)^2
+ \frac{\s}{2 v_s^2} \dot{\varphi}_i^2 \r) \text{.}
\label{eqn:A:bilayer-short-range-action}
\end{equation}
The fields $\varphi_i(\br, \tau)$ are periodic in imaginary time, and have the following Fourier decomposition:
\begin{eqnarray}
\varphi_i (\br, \tau) &=& T\sum_{\on} \int\kmeas \tilde{\varphi}_i(\bk,\on) e^{i \bk \cdot \br -i \on\tau} \text{,}\\
\tilde{\varphi}_i(\bk,\on) &=& \intt_{0}^{\beta} d\tau \int d^2 r\, \varphi_i (\br, \tau) \text{.}
	e^{-i \bk \cdot \br +i\on\tau} \nonumber
\end{eqnarray}
In Fourier space the action takes the form:
\begin{equation}
S_i=T \sum_{\on}\int\kmeas \l(\frac{\s}{2}\r) \l(\bk^2 + \frac{1}{v_s^2} \on^2 \r) |\vfk_i|^2  \text{.}
\end{equation}
Here, it is convenient to write the Josephson coupling using two auxiliary fields
$\phi(\tau)$ and $\zeta(\tau)$:
\begin{equation}
S_J =\int_0^\beta d\tau \Bigg\{ i\zeta \Big[ \phi-\big(\varphi_1(0,\tau)-\varphi_2(0,\tau)\big) \Big]
- \tc \cos (\phi)  \Bigg\}
\label{eqn:A:sj-auxfields}
\end{equation}
Integrating out $\zeta(\tau)$ identifies $\phi(\tau)$ with the phase difference across the point contact.  

It is now a simple matter to integrate out $\varphi_i$ and $\zeta$ and obtain an action only in terms of $\phi$:
\begin{equation}
	S=\frac{\s T}{4}\sum_{\on}\fk(\on)^2\l[\int\frac{d^2 k}{(2\pi)^2}
  \frac{1}{\bk^2+(\on/v_s)^2}\r]^{-1} - \tc \int_0^\beta d\tau \, \cos \phi(\tau).
\label{eqn:A:bi-sr-action-step1}
\end{equation}
In Eq.~(\ref{eqn:A:bi-sr-action-step1}), we can easily carry out the $k$ integral by imposing a short-distance cutoff for the modes in the planes, thus making the restriction
$|\bk|<k_{{\rm max}}$.  Defining $\omega_c = v_s k_{{\rm max}}$, we obtain the point-contact effective action 
\begin{equation}
S=\pi\s T\sum_{\on}|\fk(\on)|^2\l[\ln\frac{\on^2+\omega_c^2}{\on^2}\r]^{-1} - \tc
 \int_0^\beta d\tau\, \cos \phi(\tau).
 \label{eqn:bilayer-sr-action}
\end{equation}

\subsection{Bilayer geometry, Coulomb interactions}

In the case of Coulomb interactions, the action for one plane in terms of both the phase and density is:
\begin{equation}
S_i = \intt_0^\beta d\tau \Bigg\{ \int d^2 \br \Big( i \dot{\varphi}_i n_i + \frac{\s}{2} ( \nabla \varphi_i )^2 \Big)
+ \frac{(2 e)^2}{2} \int d^2 \br\, d^2 \br' \frac{n_i(\br) n_i(\br')}{| \br - \br'|} \Bigg\} \text{.}
\end{equation}
Upon performing a Fourier transform in space but not imaginary time, this becomes:
\begin{equation}
S_i =  \intt_0^\beta d\tau \int\kmeas \Big(
i \dot{\varphi}_i(\bk) n_i(-\bk) + \frac{\s |\bk|^2}{2} |\varphi_i(\bk)|^2
+ \frac{1}{2} \frac{8 \pi e^2}{|\bk|} |n_i(\bk)|^2 \Big) \text{.}
\end{equation}
Integrating out $n_i$ and taking the Fourier transform in the imaginary-time domain, we obtain an action for the phase
variable
\begin{equation}
S_i = \frac{T}{2} \sum_{\on} \int\kmeas \Big(
	\frac{|\bk| \on^2}{8 \pi e^2} + \s |\bk|^2 \Big) |\vfk_i|^2 \text{.}
\end{equation}
Proceeding as in the previous subsection, we find the effective action for $\phi$:
\begin{equation}
S = \frac{T}{4} \sum_{\on} |\fk(\on)|^2 \Big[ \int\kmeas \frac{1}{|\bk|}
	\frac{1}{(\on^2/8 \pi e^2) + \s |\bk|} \Big]^{-1}
- \tc  \int_0^\beta d\tau\, \cos \phi(\tau) \text{.}
\end{equation}
Performing the $\bk$ integral (with the restriction $|\bk| < k_{{\rm max}}$) and defining the high-frequency cutoff $\omega_c^2 = 8\pi e^2 \s k_{{\max}}$, 
the final result is:
\begin{equation}
S = \frac{\pi\s T}{2} \sum_{\on}|\fk(\on)|^2\l[\ln\frac{\on^2+\omega_c^2}{\on^2}\r]^{-1} - \tc
 \int_0^\beta d\tau \, \cos \phi(\tau) \text{.}
 \end{equation}

\subsection{Wedge geometry, short-range interactions}

Here, we consider the wedge geometry in the case of short-range interactions, where the action for a single film has the same form as Eq.~(\ref{eqn:A:bilayer-short-range-action}).  The only difference is that now, for each film, the position integration ranges only over a wedge; working with the plane polar coordinates $(r,\psi)$, this is simply the region $0 \leq \psi \leq \theta$ and $0 < r < \infty$.  Rather than going to Fourier space, we shall transform to a basis of eigenfunctions $\chi(\br)$ of the Laplace operator satisfying the Neumann boundary conditions
\begin{equation}
\partial_\psi \chi(r, \psi = 0) = \partial_\psi \chi(r, \psi = \theta) = 0 \text{.}
\end{equation}
Letting $- \nabla^2 \chi = \lambda^2 \chi$, the eigenfunctions $\chi_{\lambda n}(\br)$ are defined in Eq.~(\ref{eqn:eigenfunctions}) and are labeled by a continuous variable
$\lambda > 0$ as well as an integer $n = 0, 1, 2, \dots$.

We can define the analog of the Fourier transform of the phase field $\varphi_i$:
\begin{eqnarray}
\label{eqn:A:wedge-transform}
\varphi_i(\br, \tau)  &=& T \sum_{\on} \int_0^\infty d\lambda \sum_{n=0}^{\infty}
e^{-i \on \tau} \chi_{\lambda n} (\br)\, \vfk_i(\lambda,n,\on) \text{,} \\
\vfk_i(\lambda,n,\on) &=& \int_0^\beta d\tau \int d^2 \br e^{i \on \tau} \chi_{\lambda n}(\br)\,
	\varphi_i(\br, \tau) \text{.} \nonumber
\end{eqnarray}
In this basis the action $S_i$ takes the form
\begin{equation}
S_i = T \sum_{\on} \int_0^\infty d\lambda \sum_{n = 0}^{\infty}
	\Big(\frac{\s}{2}\Big) \big(\lambda^2 + \frac{1}{v_s^2} \on^2 \big) |\vfk_i(\lambda,n,\on)|^2 .
\end{equation}
Noting that
\begin{equation}
\varphi_i(\br = \boldsymbol{0}) = \int_0^\infty d\lambda \sqrt{\frac{\lambda}{\theta}} \vfk_i(\lambda,n = 0) \text{,}
\end{equation}
we can proceed as above to find the effective action for the phase difference across the contact:
\begin{equation}
S = \frac{\theta \s T}{4} \sum_{\on} |\fk(\on)|^2 
	\Big[ \int_0^{\lambda_{{\rm max}}} d\lambda \frac{\lambda}{\lambda^2 + (\on/v_s)^2} \Big]^{-1}
	 - \tc \int_0^\beta d\tau\, \cos \phi(\tau) \text{,}
\end{equation}
where we have introduced the short-distance cutoff $\lambda_{{\rm max}}$. Defining $\omega_c = v_s \lambda_{{\rm max}}$
and performing the $\lambda$ integration, we obtain the final result:
\begin{equation}
S = \frac{\theta\s T}{2} \sum_{\on}|\fk(\on)|^2\l[\ln\frac{\on^2+\omega_c^2}{\on^2}\r]^{-1} - \tc
 \int_0^\beta d\tau\, \cos \phi(\tau) \text{.}
 \label{eqn:wedge-sr-action}
 \end{equation}
 Notice that this result is identical to that given in Eq.~(\ref{eqn:bilayer-sr-action}), except for the factor of $(\theta / 2\pi)$, which accounts for the geometry of the wedge.
 
\subsection{Wedge geometry, Coulomb interactions}
\label{sec:lr-wedge}

Here, we replace the Coulomb interaction with the new potential $V(\br, \br')$ discussed in Sec.~\ref{sec:formulation}.  The action for one plane, in terms of the phase and density, is
\begin{equation}
S_i = \intt_0^\beta d\tau \Bigg\{ \int d^2 \br \Big( i \dot{\varphi}_i n_i + \frac{\s}{2} ( \nabla \varphi_i )^2 \Big)
+ \frac{1}{2} \int d^2 \br\, d^2 \br' V(\br,\br') n_i(\br) n_i(\br') \Bigg\} \text{.}
\end{equation}
Making the transformation of Eq.~(\ref{eqn:A:wedge-transform}), this action becomes
\begin{equation}
S_i = T \sum_{\on} \sum_{n=0}^\infty \intt_0^\infty d\lambda \Bigg\{
\on \vfk_i(\lambda,n,\on) \tilde{n}_i(\lambda,n,-\on)
+ \frac{\s \lambda^2}{2} |\vfk_i(\lambda,n,\on)|^2
+ \frac{4 \pi e^2}{\lambda} |\tilde{n}_i(\lambda,n,\on)|^2 \Bigg\} \text{.}
\end{equation}
Upon integrating out the density, we obtain
\begin{equation}
S_i = \frac{T}{2} \sum_{\on} \sum_{n=0}^\infty \intt_0^\infty d\lambda \Big(
\s \lambda^2 + \frac{\lambda \on^2}{8 \pi e^2} \Big) |\vfk_i(\lambda,n,\on)|^2 .
\end{equation}
Proceeding as above, we find the effective action at the contact:
\begin{equation}
S = \frac{\theta \s T}{4} \sum_{\on} |\fk(\on)|^2
\Big[ \int_0^{\lambda_{{\rm max}}} d\lambda \frac{\lambda}{\lambda^2 + (\lambda \on^2 / 8 \pi e^2 \s)}
\Big]
- \tc \int_0^\beta d\tau \, \cos \phi(\tau).
\end{equation}
The $\lambda$ integral is easily evaluated, and, defining $\omega_c^2 = 8 \pi e^2 \s \lambda_{{\rm max}}$, we thus obtain the final result for the effective action:
\begin{equation}
S = \frac{\theta \s T}{4} \sum_{\on} |\fk(\on)|^2\l[\ln\frac{\on^2+\omega_c^2}{\on^2}\r]^{-1} - \tc
 \int_0^\beta d\tau\, \cos \phi(\tau) \text{.}
 \end{equation}

\section{Analyzing the imaginary-time action}
\label{app:imag-time}

\subsection{Analytic structure of the imaginary time action} 

The action (\ref{eqn:gaussian-phi-action}) for the phase difference across the contact has a rather peculiar form -- the Gaussian part is logarithmic in
frequency. In order to extract the correlation functions of the field
$\phi(t)$ via analytic continuation, we need to understand the
analytic properties of this logarithm. 

The function $\ln\frac{\on^2+\omega_c^2}{\on^2}$ can be broken into two
parts:
\be
\ln\frac{\on^2+\omega_c^2}{\on^2}=\ln\frac{\on+i\omega_c}{\on}+\ln\frac{\on-i\omega_c}{\on}.
\label{eqn:A:log-breakup}
\ee
On the right hand side of Eq.~(\ref{eqn:A:log-breakup}), the first term has a branch cut stretching
between $\on=0$ and $\on=-i\omega_c$; similarly, the second term has a
branch cut between $\on=0$ and $\on=i\omega_c$. As one may expect, in
order to be able to find correlation functions, we
need to resolve the values of the logarithm functions in the vicinity
of their branch cuts. The results are:
\begin{subequations}
\begin{eqnarray}
\l.\ln\frac{\on^2+\omega_c^2}{\on^2}\r|_{\o=\epsilon+ix} &=& \ln\frac{\omega_c^2-x^2}{x^2}-i\pi\\
\l.\ln\frac{\on^2+\omega_c^2}{\on^2}\r|_{\o=-\epsilon+ix} &=& \ln\frac{\omega_c^2-x^2}{x^2}+i\pi \\
\l.\ln\frac{\on^2+\omega_c^2}{\on^2}\r|_{\on=-\epsilon-ix} &=& \ln\frac{\omega_c^2-x^2}{x^2}-i\pi\\
\l.\ln\frac{\on^2+\omega_c^2}{\on^2}\r|_{\on=\epsilon-ix} &=& \ln\frac{\omega_c^2-x^2}{x^2}+i\pi
\label{eqn:A:branch-cut-resolution}
\end{eqnarray}
\end{subequations}
where $\epsilon>0$ is an infinitesimal real number indicating which side of
the branch cut we refer to, and $0<x<\omega_c$. The properties of
$\ln\frac{\on^2+\omega_c^2}{\on^2}$ in the complex plane are shown in
Fig.~\ref{fig:A:log-analytic-properties}.

\begin{figure}
\includegraphics[width=8cm]{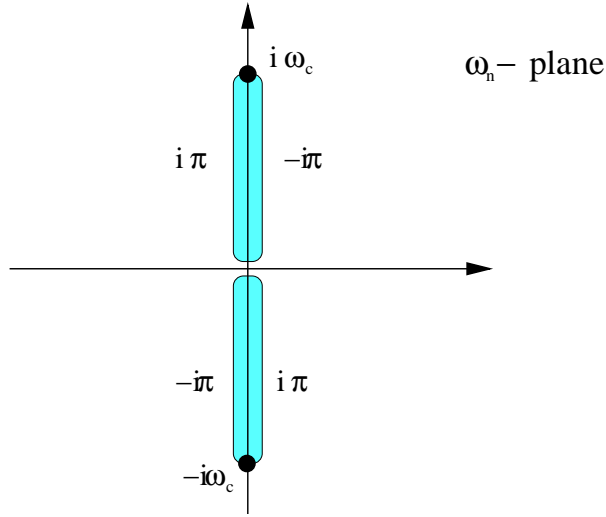}
\caption{Analytic properties of $\ln\frac{\on^2+\omega_c^2}{\on^2}$. The
  function has two branch cuts originating at $\on=0$ and terminating
  at $\on=\pm i\omega_c$. The values $\pm i\pi$ resolve the value of the
  function near the branch cut, and are added to
  $\ln\frac{\omega_c^2-|\o|^2}{|\o|^2}$ to give the value of the
  function there.}
\label{fig:A:log-analytic-properties}
\end{figure}

\subsection{Phase correlation function}
\label{app:phase-corr-fn}

In the weak-coupling limit, the equilibrium correlations of $\phi(t)$ are related to the linear response conductance by a Kubo formula as described in Sec.~\ref{sec:weak-coupling}. We are now in a position to find these correlations from the imaginary-time action, Eq.~(\ref{eqn:gaussian-phi-action}).

\begin{figure}
\includegraphics[width=8cm]{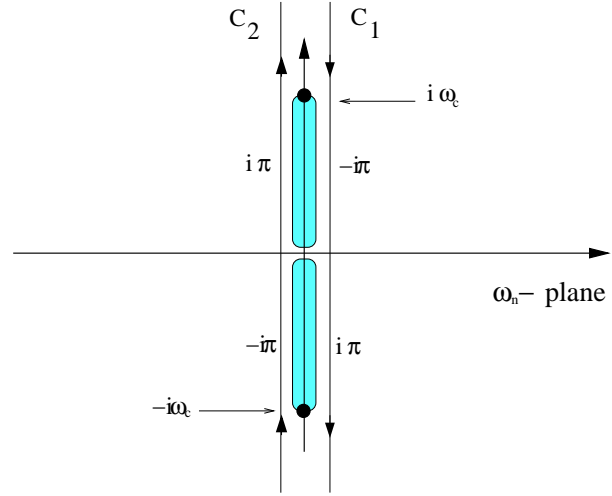}
\caption{Contours of integration in the $\on$-plane for the Matsubara frequency sum
  Eq. (\ref{eqn:A:mf-sum}).  $C_1$ extends from $\epsilon + i \infty$ to $\epsilon - i \infty$, and $C_2$ ranges from $-\epsilon - i\infty$ to $-\epsilon + i \infty$. $C_1$ and $C_2$ are closed by semicircles at infinity (not shown) in the right and left half-planes, respectively. The poles of  $\l(1-e^{-i\on\beta}\r)^{-1}$ lie on the real axis, and are contained within $C_1$ and $C_2$, whereas the
  branch cuts of the logarithm function lie outside $C_1$ and $C_2$.}
  \label{figA2}
\end{figure}

For $\tc = 0$, the imaginary-time phase correlator is
\begin{equation}
\langle\phi(\tau)\phi(0) - \phi^2 \rangle=T\sum_{\on \neq 0}\frac{\pi^2}{4 c \s}
\ln \Big[ \frac{\on^2+\omega_c^2}{\on^2} \Big] (e^{-i\on\tau} - 1).
\label{A3.1}
\end{equation}
As usual, we can rewrite the sum over Matsubara
frequencies as an integral. This is done as follows:
\be
T\summ_{\on\neq 0}\frac{\pi^2}{4 c \s}
\ln\Big[\frac{\on^2+\omega_c^2}{\on^2}\Big] (e^{-i\on\tau} - 1)=
\frac{\pi^2}{4 c \s} \intt_{C_1+C_2}\frac{d\on}{2\pi}\frac{e^{-i\on\tau} - 1}{1-e^{-i\on\beta}}\ln\frac{\on^2+\omega_c^2}{\on^2}.
\label{eqn:A:mf-sum}
\ee
In Eq.~(\ref{eqn:A:mf-sum}) we introduce the contours $C_1$ and $C_2$ such
that we include the poles of $\l(1-e^{-i\on\beta}\r)^{-1}$ but not the
branch cuts of the logarithm function; these contours are shown in
Fig.~(\ref{figA2}). The contours run infinitesimally to either side of the imaginary axis and are closed by semicircles at infinity.  The only nonzero contributions arise near the imaginary axis for $|\on| < \omega_c$, and we thus have
\be
\langle\phi(\tau)\phi(0) - \phi^2 \rangle=\frac{\pi^2}{4 c \s} \l[\intt_{\epsilon+i\omega_c}^{\epsilon-i\omega_c}+\intt_{-\epsilon-i\omega_c}^{-\epsilon+i\omega_c}\r]\frac{d\on}{2\pi}\frac{e^{-i\on\tau} - 1}{1-e^{-i\on\beta}}\ln\frac{\on^2+\omega_c^2}{\on^2} \text{,}
\label{A3.3}
\ee
where $\epsilon=0^+$. These integrals can be recast as integrals on the real
axis by using the branch-cut resolution of Eq.~(\ref{eqn:A:branch-cut-resolution}):
\begin{eqnarray}
\langle\phi(\tau)\phi(0) - \phi^2 \rangle&=& 
\frac{\pi^2}{4 c \s}
\l(-\intt_{0}^{\omega_c}\frac{idx}{2\pi}\frac{e^{x\tau} - 1}{1-e^{x\beta}}\l(\ln\frac{\omega_c^2-x^2}{x^2}-i\pi\r)\r.
+\intt_{0}^{\omega_c}\frac{idx}{2\pi}\frac{e^{x\tau} - 1}{1-e^{x\beta}}\l(\ln\frac{\omega_c^2-x^2}{x^2}+i\pi\r) \label{A3.4} \\
&+& \intt_{0}^{\omega_c}\frac{idx}{2\pi}\frac{e^{-x\tau} - 1}{1-e^{-x\beta}}\l(\ln\frac{\omega_c^2-x^2}{x^2}-i\pi\r)
\l.-\intt_{0}^{\omega_c}\frac{idx}{2\pi}\frac{e^{-x\tau} - 1}{1-e^{-x\beta}}\l(\ln\frac{\omega_c^2-x^2}{x^2}+i\pi\r)\r) \text{.} \nonumber
\end{eqnarray}
Combining terms, we obtain
\begin{equation}
\langle\phi(\tau)\phi(0) - \phi^2 \rangle
=\frac{\pi^2}{4 c \s} \intt_0^{\omega_c} dx\l((\cosh(x\tau)-1)\coth\l(\frac{\beta
  x}{2}\r)-\sinh(x\tau)\r) \text{.}
\label{A3.6}
\end{equation}
The last step is to analytically continue this expression
to real time, $\tau\rightarrow it$, which yields
\begin{equation}
\langle\phi(t)\phi(0)-\phi(0)^2\rangle=\frac{\pi^2}{4 c \s} \intt_0^{\omega_c}dx\l(\l(\cos(xt)-1\r)\coth\l(\frac{\beta
  x}{2}\r)-i\sin(xt)\r).
\label{A3.7}
\end{equation}
We note that we have also obtained this result via the Hamiltonian formulation, working directly in real time.

\subsection{Dual correlation function}
\label{app:imag-time-dual-cf}

In Sec.~\ref{sec:duality}, we derived the dual form of the action at the point contact in terms of the charge-imbalance variable $\theta(\tau)$.  In order to calculate the resistance, as discussed in Sec.~\ref{sec:resistance-calculation}, we need the correlation function $\langle \hat{\theta}(t) \hat{\theta}(0) - \hat{\theta}^2 \rangle_{t_v = 0}$.  This can be obtained by analytic continuation from its imaginary-time analog:
\begin{equation}
\langle \theta(\tau) \theta(0) - \theta^2 \rangle =
T \sum_{\on} \frac{\bar{\tc}}{1 + \frac{\alpha \bar{\tc}}{\theta \s} \ln \frac{\on^2 + \omega_c^2}{\on^2}}
\Bigg( \frac{e^{-i \on \tau} - 1}{\on^2} \Bigg) .
\end{equation}

We can proceed exactly as we have just done for the phase correlation function.  Rewriting the Matsubara sum as an integral over the contours $C_1$ and $C_2$ of Fig.~\ref{figA2}, the result is:
\begin{eqnarray}
\langle \theta(\tau) \theta(0) - \theta^2 \rangle &=&
\int_{C_1 + C_2} \frac{ d \on}{2\pi} \frac{e^{-i \on \tau} -1}{\on^2 (1 - e^{-i \on \beta})}
\frac{\bar{\tc}}{1 + \frac{\alpha \bar{\tc}}{\theta \s} \ln \frac{\on^2 + \omega_c^2}{\on^2}} \\
&=& \frac{\theta \s}{\alpha} \int_0^{\omega_c} \frac{dx}{x^2}
\Bigg( \big(\cosh(\tau x) - 1 \big) \coth \Big( \frac{\beta x}{2} \Big) - \sinh (\tau x) \Bigg)
\frac{1}{\pi^2 + \Big( \frac{\theta\s}{\alpha \bar{\tc}} + \ln \frac{\omega_c^2 - x^2}{x^2} \Big)^2} \text{,}
\end{eqnarray}
where the second equality is obtained by manipulations analogous to those of Appendix~\ref{app:phase-corr-fn}.  To analytically continue, we replace $\tau \to i t$ and obtain the sought correlator:
\begin{equation}
\langle \hat{\theta}(t) \hat{\theta}(0) - \hat{\theta}^2 \rangle =
\frac{\theta\s}{\alpha} \int_0^{\omega_c} \frac{d x}{x^2} 
\Bigg( \big(\cos(x t) - 1 \big) \coth \Big( \frac{\beta x}{2} \Big) - i \sin (x t) \Bigg)
\frac{1}{\pi^2 + \Big( \frac{\theta\s}{\alpha \bar{\tc}} + \ln \frac{\omega_c^2 - x^2}{x^2} \Big)^2} \text{.}
\end{equation}

\section{Evaluation of resistance integrals}
\label{app:integrals}

%Evaluation of $f(0)$ and $f''(0)$, criterion for validity of the saddle point evaluation.
\subsection{Low-temperature form of $f(0)$ and $f''(0)$}
\label{app:integrals-lowT-limit}

Here, we derive the low-temperature forms of $f(0)$ and $f''(0)$, which go into the saddle-point evaluation of the tunneling resistance [see Eq.~\ref{eqn:saddle-point-resistance}].  Recalling the definition of $f(t)$ in Eq.~(\ref{eqn:defn-of-f-function}), it is helpful to express it in the form
\begin{equation}
f(t) = \frac{8 c \s}{T}  \int_0^{\beta \omega_c / 2} \frac{du}{u^2}
\frac{\cos(2 u t / \beta) - \cosh(u) }{\sinh(u)}
\frac{1}{\pi^2 + \big[ 4 c \s  / \pi^2 \bar{\tc} + \ln \big((\beta\omega_c / 2 u)^2 - 1\big) \big]^2} \text{.}
\label{eqn:app:dimensionless-foft}
\end{equation}

First consider $f(0)$, which we write as
 \begin{equation}
 f(0) = - \frac{8 c \s}{T} \big(f_1(T) + f_2(T) \big) \text{,}
 \end{equation}
 where
\begin{equation}
f_1(T) = \int_0^1 \frac{du}{u^2} \frac{\cosh(u) - 1}{\sinh u} 
\frac{1}{\pi^2 + \big[4 c \s / \pi^2 \bar{\tc} + \ln \big(  (\beta\omega_c/ 2 u)^2 -1\big)\big]^2}
\text{,}
\end{equation}
and $f_2(T)$ is of the same form, but evaluated between the limits $1$ and $\beta \omega_c /2$.
%\begin{equation}
%f_2(T) = \int_1^{\beta\omega_c/2} \frac{du}{u^2} \frac{\cosh(u) - 1}{\sinh u} 
% \frac{1}{\pi^2 + \big( \frac{\theta\s}{\alpha\bar{\tc}} + \ln \big(\frac{(\beta\omega_c/2)^2 - u^2}{u^2}\big)\big)^2}
% \text{.}\label{eqn:f2-of-T}
 %\end{equation}

To evaluate $f_1(T)$, we first note that, to very good accuracy, $\ln \big((\beta\omega_c/ 2 u)^2 - 1\big) \approx 2\ln(\beta\omega_c / 2 u)$ for $u \in [0,1]$.  We proceed by expanding $[\cosh(u) - 1]/ \sinh(u)$ in powers of $u$.  Keeping only the leading term (at small $u$), we have
\begin{eqnarray}
f_1(T) &\approx& \frac{1}{8} \int_0^1 \frac{du}{u} 
\frac{1}{(\pi/2)^2 + \big[2 c \s / \pi^2 \bar{\tc} + \ln \big(  \beta\omega_c/ 2 u \big)\big]^2} \nonumber\\
&=& \frac{1}{8} \Big\{ 1 -  \frac{2}{\pi} \operatorname{arctan} \Big[ \frac{2}{\pi} \Big( \ln (\omega_c / 2 T)
+ 2 c \s / \pi^2 \bar{\tc} \Big) \Big] \Big\} \text{.}
\end{eqnarray}
When the argument of the arctangent is large, 
\begin{equation}
f_1(T) \approx \frac{1}{8} \frac{1}{\ln(\omega_c / 2 T) + 2 c \s / \pi^2 \bar{\tc} } \,\, \text{.}
\label{eqn:f1ofT}
\end{equation}
This holds within $1\%$, as long as the argument of the arctangent is above about $4$, which is always easily satisfied for reasonable parameters.
The higher terms in the power-series expansion fall off as $1/\ln^2(\omega_c /T)$ or faster in the $T \to 0$ limit, so the $T \to 0$ asymptotic behavior of $f_1(T)$ comes from Eq.~(\ref{eqn:f1ofT}), and reads
\begin{equation}
f_1(T) \sim \frac{1}{8} \frac{1}{\ln (\omega_c / 2 T)} \text{.}
\end{equation}

Now consider $f_2(T)$ -- we will show that it is dominated by $f_1(T)$ as $T \to 0$.  As
$[\cosh(u) -1]/\sinh(u) < 1$ for all $u \in [1, \beta \omega_c /2]$, we have
\begin{equation}
f_2(T) < \int_1^{\beta\omega_c/2} \frac{du}{u^2} 
\frac{1}{\pi^2 + \big[ 4 c \s / \pi^2 \bar{\tc} + \ln \big((\beta\omega_c/2 u)^2 - 1\big)\big]^2} \text{.}
\end{equation}
We can break this integral into two more pieces, writing $f_2(T) < g_A(T) + g_B(T)$, where
\begin{equation}
g_A(T) = 
\int_1^{\epsilon \sqrt{\beta\omega_c}/2} \frac{du}{u^2} 
\frac{1}{\pi^2 + \big[ 4 c \s  / \pi^2 \bar{\tc} + \ln \big((\beta\omega_c/2 u)^2 - 1\big)\big]^2} \text{,}
\end{equation}
and $g_B(T)$ takes the same form but is integrated from $\epsilon \sqrt{\beta\omega_c}/2$
to $\beta\omega_c/2$.  We take the low-temperature limit of $g_A + g_B$, keeping $\epsilon$ fixed.  

As long as $\epsilon$ is not too large,  we have
\begin{eqnarray}
g_A(T) &<& \frac{1}{\pi^2 +\big[4 c \s / \pi^2 \bar{\tc} + \ln \big( (\sqrt{\beta\omega_c}/ \epsilon)^2 -1 \big)\big]^2} 
\int_1^{\epsilon \sqrt{\beta\omega_c}/2} \frac{du}{u^2} \\
&\sim& \frac{1}{\ln^2 (\omega_c / T)} \text{.}
\end{eqnarray}
On the other hand,
\begin{equation}
g_B(T) < \frac{1}{\pi^2} \int_{\epsilon \sqrt{\beta\omega_c}/2}^{\beta\omega_c/2} \frac{du}{u^2}
\sim \frac{1}{\epsilon \pi^2} \sqrt{\frac{T}{\omega_c}} \text{,}
\end{equation}
which is dominated by $g_A$.  Therefore, we have
\begin{equation}
f_2(T) \lesssim \frac{1}{\ln^2 (\omega_c / T) } \text{,}
\end{equation}
as claimed.  
Thus we may write
\begin{equation}
f(0) \approx - \frac{c \s}{T} \frac{1}{\ln (\omega_c / 2 T) + 2 c \s / \pi^2 \bar{\tc}} \,\, \text{.}
\label{eqn:approx-f0-formula}
\end{equation}
%See Appendix~\ref{app:integrals-numerical} for a discussion of the quantitative accuracy for this formula.
We also have the low-temperature asymptotic form
\begin{equation}
f(0) \sim - \frac{c \s}{T \ln( \omega_c / 2 T) } \text{.}
\end{equation}

Now we evaluate $f''(0)$, by writing
\begin{eqnarray}
f''(0) &=& - 16 c \s T \int_0^{\beta\omega_c/2} \frac{du}{\sinh u}
\frac{1}{\pi^2 + \big[ 4 c \s / \pi^2 \bar{\tc} + \ln \big( (\beta\omega_c / 2 u)^2 - 1 \big) \big]^2} \nonumber\\
&=& -16 c \s T \Big( \int_0^1 du + \int_1^{\beta\omega_c/2} du \Big)
\frac{1}{\sinh u} 
\frac{1}{\pi^2 + \big[ 4 c \s / \pi^2 \bar{\tc} + \ln \big( (\beta\omega_c / 2 u)^2 - 1 \big) \big]^2} \nonumber \\
&\equiv& - 16 c \s T (\tilde{f}_1(T) + \tilde{f}_2(T)) \text{.}
\end{eqnarray}
The function $\tilde{f}_1(T)$ can be evaluated similarly to $f_1(T)$, by writing $1/\sinh u = (1/u)(u/\sinh u)$, and
expanding $u / \sinh u$ in powers of $u$.  The leading term again gives the dominant $T \to 0$ behavior, and proceeding as above we find
\begin{equation}
\tilde{f}_1(T) \approx \frac{1}{4} \frac{1}{\ln(\omega_c / 2 T) + 2 c \s / \pi^2 \bar{\tc}} \,\, \text{.}
\end{equation}
Furthermore, as $1/\sinh u$ decays exponentially at large $u$, it is clear that $\tilde{f}_2(T)$ will go as $1/\ln^2 (\omega_c /T)$ at low temperature, and will be dominated by $\tilde{f}_1(T)$.
Therefore, we have
\begin{equation}
f''(0) \approx -8 c \s T \frac{1}{\ln(\omega_c / 2 T) + 2 c \s / \pi^2 \bar{\tc}} \,\, \text{,}
\label{eqn:app-f0pp-expression}
\end{equation}
and, at low-temperature, 
\begin{equation}
f''(0) \sim - \frac{8 c \s T}{\ln (\omega_c / 2 T)} \text{.}
\end{equation}

\subsection{Evaluation of $f(0)$ and $f''(0)$ for large and small $\tc / \s$}
\label{app:integrals-poor-contact}

The analysis above yielded the low-temperature asymptotic forms for $f(0)$ and $f''(0)$ valid for $T \ll T_J$, and hence the universal low-temperature limit of the resistance.  Furthermore, we obtained approximate formulas for these quantities that reduce to the asymptotic forms for $T \ll T_J$, and include some dependence on $\bar{\tc}$ away from this limit.  Here, we show that these formulas are exact, not only at very low temperatures, but also in suitably defined good- and poor-contact limits.

The good-contact limit is simply $\tc \gg \s$.  Taking $\tc / \s \to \infty$, it is clear that all dependence on $\bar{\tc}$ simply drops out of $f(t)$, and at low temperatures one can follow the analysis of the previous subsection to obtain the universal forms for $f(0)$ and $f''(0)$.  The only requirement on the temperature in this case is that $\s / T \gg 1$, so that $E_A(T)/ T$ is large, and terms of higher-order in $t_v$ do not contribute appreciably to the resistance.

The poor-contact limit we must consider here is not merely the ``na\"{\i}ve'' poor contact limit of 
$\tc \ll \s$.  Because the resistance crosses over to the good-contact form at low enough temperature, we must also demand that $T \gg T_J$.  With these conditions, the approximate forms of the previous subsection become:
\begin{equation}
f(0) \approx - \frac{\pi^2 \bar{\tc}}{2 T} \text{,}
\end{equation}
and
\begin{equation}
f''(0) \approx - 4 \pi^2 \bar{\tc} T \text{.}
\end{equation}
These results are actually asymptotically exact for $\tc \ll \s$ and $T \gg T_J$.  Furthermore, we should require that $\tc / T \gg 1$, so that the resistance is dominated by the contribution of leading-order 
in $t_v$.  

To see this, let us revisit the analysis of $f(0)$ given in Appendix~\ref{app:integrals-lowT-limit}, and consider the functions $f_1(T)$ and $f_2(T)$.  
Again, $f_1(T)$ can be analyzed by expanding $(\cosh(u) - 1)/\sinh u$ in powers of $u$.  Only the leading term in this expansion leads to a contribution to $f_1(T)$ of order $\bar{\tc} / \s$ -- this is exactly the integral we already evaluated to arrive  at the approximate formula, Eq.~(\ref{eqn:approx-f0-formula}).  All other terms in the power-series expansion give contributions of order $( \bar{\tc} / \s)^2$.  Similarly, it is straightforward to show that $f_2(T) \sim (\bar{\tc} / \s)^2$.  Therefore, we can write:
\begin{equation}
f(0) \approx - \frac{\pi^2 \bar{\tc}}{2 T} \frac{1}{1 + \frac{\pi^2 \bar{\tc}}{2 c \s} \ln(\omega_c/ 2 T)} \text{.}
\end{equation}
Note that it is important to require that $T \gg T_J$; otherwise, $(\bar{\tc}/ \s) \ln(\omega_c / 2 T)$ would not be small, and it would no longer be legitimate to neglect higher-order terms in $\bar{\tc} / \s$.  However, for $T \gg T_J$ we have:
\begin{equation}
f(0) \sim - \frac{\pi^2 \bar{\tc}}{2 T} \text{.}
\end{equation}

Similarly, it can be shown that, in the same limit,
\begin{equation}
f''(0) \sim -4 \pi^2 \bar{\tc} T \text{.}
\end{equation}

\subsection{Validity of the saddle-point approximation}
\label{app:integrals-saddle-pt-validity}

We here derive a \emph{conservative\/} criterion for the validity of the saddle-point evaluation of the resistance integral, Eq.~(\ref{eqn:saddle-point-resistance}).  This criterion is simply that, in the power-series expansion of $f(t)$, terms higher than 2nd order in $t$ should be small when the 2nd order term is of order unity.  Formally, we require that
\begin{equation}
\frac{1}{(2n)!} | f^{(2n)}(0)  | t_0^{2n} \ll 1\text{,}
\label{eqn:saddle-pt-criterion}
\end{equation}
where $n > 1$, and $t_0$ is defined by
\begin{equation}
\frac{1}{2} | f''(0)| t_0^2 = 1 \text{.}
\label{eqn:t0-condition}
\end{equation}
The
higher-order terms alternate in sign, and will thus tend to cancel one another out, so we expect the saddle-point approximation to be even better than the results derived below suggest.

First, we must estimate $f^{(2n)}(0)$, which we write in the form:
\begin{equation}
f^{(2n)}(0) = 16 c \s (-1)^n (2T)^{2n -1} I_{2n} \text{,}
\end{equation}
where we have defined
\begin{equation}
I_{2n} \equiv \int_0^{\beta\omega_c/2} \frac{u^{2(n-1)}}{\sinh u}
\frac{1}{\pi^2 + \big[ 4 c \s / \pi^2 \bar{\tc} + \ln \big( (\beta\omega_c/2u)^2 - 1 \big) \big]^2} \text{.}
\end{equation}
As the logarithm varies very slowly, and the integrand decays exponentially at large $u$, we can approximate as follows:
\begin{eqnarray}
I_{2n} &\approx& \frac{1}{\pi^2 + \big[ 4 c \s / \pi^2 \bar{\tc} + 2 \ln \big( \beta \omega_c / 2 u_{{\rm max}} \big) \big]^2} \int_0^{\infty} \frac{u^{2(n-1)}}{\sinh u} \nonumber \\
&=&  \frac{1}{\pi^2 + \big[ 4 c \s / \pi^2 \bar{\tc} + 2 \ln \big( \beta \omega_c / 2 u_{{\rm max}} \big) \big]^2}
\frac{(4^n - 2) (2n - 2)! \,\zeta(2n - 1)}{2^{2n-1}} \text{.}
\end{eqnarray}
Here, $u_{{\rm max}}$ is the value of $u$ at which $u^{2(n-1)}/\sinh u$ is maximum;  $u_{{\rm max}} \approx 2(n-1)$, to good accuracy.  Furthermore, we can safely make the approximations 
$4^n - 2 \approx 4^n$ and $\zeta(2n-1) \approx 1$, so that
\begin{equation}
\frac{1}{(2n)!} I_{2n} \approx \frac{1}{n(2n-1)} 
\frac{1}{\pi^2 + \big[ 4 c \s / \pi^2 \bar{\tc} + 2 \ln \big( \beta \omega_c / 4 (n-1) \big) \big]^2} \text{.}
\end{equation}

Then, using Eqs.~(\ref{eqn:t0-condition}) and~(\ref{eqn:app-f0pp-expression}) to determine $t_0$, we have
\begin{eqnarray}
\frac{1}{(2n)!} | f^{(2n)}(0) | t_0^{2n} &\approx&
\frac{8 c \s}{n(2n-1) T \Big[ \pi^2 + \big[4 c \s / \pi^2 \bar{\tc} + 2 \ln \big( \omega_c / 4 T (n-1) \big) \big]^2 \Big]}
\Bigg( \frac{T \big[ \ln(\omega_c / 2T) + 2 c \s / \pi^2 \bar{\tc} \big]}{c \s} \Bigg)^n \nonumber \\
&\approx&
\frac{8 \big( \ln(\omega_c / 2 T) + 2 c \s / \pi^2 \bar{\tc} \big)}{n(2n-1)
\Big[ \pi^2 + \big[ 4 c \s / \pi^2 \bar{\tc} + 2 \ln \big( \omega_c / 4 T (n-1) \big) \big]^2 \Big]}
\Big| f(0) \Big|^{1-n}  \text{.}
\end{eqnarray}
The important part of this expression is the factor of $|f(0)|^{1-n}$ on the right hand side.
We see that the condition Eq.~(\ref{eqn:saddle-pt-criterion}) for the validity of the saddle-point approximation is simply that $|f(0)|$ is large
or, equivalently, that the argument of the exponential in the resistance formula is large and negative.

\section{Variational Method}
\label{app:variational}

The main result of this paper is Eq.~(\ref{eqn:resistance}) for the contact resistance. 
This result, which was obtained by starting from the good-contact limit and
implementing a perturbation theory in phase-slip events, is expected
to apply for a large range of parameters that stretches all the way to
$\tc \sim T$ at low temperatures.
  
As discussed in the Introduction and Sec.~\ref{sec:formulation},
several pieces of evidence justify this expectation.  Here, we provide
further supporting evidence, obtained from a variational approach of
the type first employed by Feynman \cite{Feynman-Hibbs}.

In the context of Josephson or point-contact systems, the variational approach used here allows
us to find a quadratic substitute to the cosine terms in the imaginary-time action.  This method was
used to find the phase diagram of the resistively shunted
Josephson junction,\cite{fisher85} and has since been used in several
other contexts, \emph{e.g.},~quasiparticle dissipation
\cite{schoen90}.  Although this method is an uncontrolled approximation, it may provide some information, if interpreted appropriately.

\begin{figure}
\includegraphics[width=8cm]{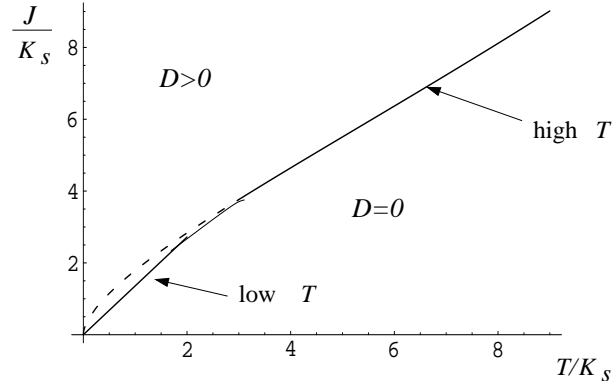}
\caption{Phase diagram resulting from the variational approach. A
  nonzero self-consistent solution to Eq.~(\ref{v.9}) for the variational parameter $D$, exists 
  for $\tc>J_0(T)$, given by Eq.~(\ref{v.10}). The dashed line is
  the continuation of an high-$T$ expression (only the explicit solution of the
  low-T regime is discussed in the text).   \label{varfig}}
\end{figure}
 
The starting point of the variational calculation is the action in
Eqs. (\ref{eqn:gaussian-phi-action}) and (\ref{eqn:josephson-coupling}), but written in terms of an Ansatz quadratic action
plus a perturbation:
\be
S=S_0+S_1,
\label{v.1}
\ee
with
\begin{equation}
S_0 = \frac{2 c \s T}{\pi^2}
\summ_{\o}\vfk(\o)^2\l[\ln\frac{\o^2+\omega_c^2}{\o^2}\r]^{-1}+\int
d\tau \frac{1}{2}D\vf(\tau)^2=T\summ_{\o}\vfk(\o)^2 g^{-1}(\o) \text{,} \label{eqn:varaction}
\end{equation}
and
\begin{equation}
S_1 = \int d\tau \l[\tc\l(1-\cos \vf(\tau)\r)-\frac{1}{2}D\vf(\tau)^2\r] \text{.}
\label{v.2}
\end{equation}
The variational parameter is $D$, which is the coefficient of the $\varphi^2$ ``mass'' term in the quadratic Ansatz action $S_0$.

To find the best $S_0$ to describe the point-contact system, we follow
the Feynman strategy by minimizing the free energy with respect to
$D$:
\be
\tilde{F}= -\ln Z_0+\l\langle S_1 \r\rangle _{S_0}= -\ln\int
D[\vf]\exp\l(-S_0\r)+\frac{\int D[\vf]\exp\l(-S_0\r)S_1}{\int
  D[\vf]\exp\l(-S_0\r)}.
\label{v.3}
\ee
A straightforward calculation yields
\be
\frac{\partial \tilde{F}}{\partial D}=\l\langle S_1\cdot \frac{\partial S_1}{\partial D}\r\rangle_{S_0} - \l\langle S_1 \r\rangle_{S_0}\l\langle\frac{\partial S_1}{\partial D} \r\rangle_{S_0},
\label{v.4}
\ee
where $\l\langle\ldots\r\rangle_{S_0}$ denotes the thermal average with
respect to the action $S_0$. This leads to the following
self-consistency condition for $D$:
\be
D=\tc \exp\l[-\frac{1}{2}T\summ_{\o}g(\o)\r].
\label{v.5}
\ee
To evaluate the sum $T\summ_{\o}g(\o)$, we follow
exactly the same steps as in Appendix \ref{app:imag-time}, and thus obtain
\be
T\summ_{\o}g(\o)= \frac{1}{\pi} \intt_0^{\omega_c}\frac{dx}{2\pi}\coth\l(\frac{x\beta}{2}\r)\frac{4 c \s}{\l(\pi
  D\r)^2 +\l(\frac{2 c \s}{\pi^2} +D\ln\frac{\omega_c^2-x^2}{x^2}\r)^2}.
\label{v.7}
\ee
The dominant contribution to the above integral comes from the region $x<2T$,
where the hyperbolic cotangent is large. We therefore make the 
approximations: $\coth x\beta/2\approx 2/(x\beta)$ and $\ln \omega_c/x
\gg \pi$. This allows us to evaluate the integral approximately,
\be
T\summ_{\o}g(\o)\approx \frac{4 c \s T}{\pi^2} \intt_0^{2T} \frac{dx}{x}
\frac{1}{\l(\frac{2 c \s}{\pi^2}+2D\ln \frac{\omega_c}{x}\r)^2}=\frac{2 c \s
  T}{\pi^2 D}\frac{1}{2D\ln \frac{\omega_c}{2T}+ \frac{2 c \s}{\pi^2}},
\label{v.8}
\ee
and hence arrive at a transcendental self-consistency condition for
$D$:
\be
\tc\exp\l[-\frac{c \s T}{\pi^2 D}\frac{1}{2D\ln
    \frac{\omega_c}{2T}+\frac{2 c \s}{\pi^2}}\r]=D.
\label{v.9}
\ee

For all values of the parameters this condition has the trivial
solution $D=0$. It also has a $D \neq 0$ solution for all temperatures,
but only for $\tc>J_0(T)$; i.e., the coupling must be stronger then a temperature-dependent critical
value, $J_0(T)$ (see Fig. \ref{varfig}).  Because we are interested only in temperatures well below $T_{{\rm BKT}}$, we focus on the low-temperature limit of the self-consistency relation, which is specified by 
$T \ln (\omega_c / 2 T) \ll 2\pi \s$.  In this limit, we 
find that $J_0(T)$ has the following
form:
\begin{equation}
J_0(T)\approx \frac{e}{2} T \text{,}
\label{v.10}
\end{equation}
where $e$ is the base of the natural logarithm, \emph{not\/} the electron charge.
According to Eq.~(\ref{v.9}), $D$ jumps discontinuously at the
critical line $J_0(T)$, from $0$ for $\tc<J_0(T)$ to
\begin{equation}
D_0 \approx \frac{\tc}{e}
\label{v.11}
\end{equation}
for $\tc > J_0(T)$.

The appearance of a self-consistent mass term suggests that 
the Josephson coupling $\tc$ should be treated as large,
because the point-contact phase difference tends to localize in one of
the wells of the Josephson potential.  Therefore, the proper approach in
the regime $\tc>J_0(T)$ is to expand about the good-contact limit, where
tunneling between minima of the Josephson potential is
treated perturbatively.  Indeed, the region where we find a self-consistent nonzero
mass (Eq. \ref{v.10}) essentially coincides with the
regime of validity of the resistance formula (\ref{eqn:resistance}), which is roughly given by
\be
T \lesssim E_A(T) \text{.}
\ee 
This provides further justification that
the expansion about the good-contact limit is valid at low temperature, even for
small Josephson coupling.

\bibliography{points}

\end{document}